\documentclass[final]{siamltex}
\usepackage{amssymb}
\usepackage{showkeys}
\usepackage{graphicx}
\usepackage{amsmath}
\usepackage{graphicx}
\usepackage{epsfig}
\usepackage{labelfig}
\usepackage{amsmath}
\usepackage{color}
\usepackage{afterpage}
\usepackage{verbatim}
\usepackage{umlaut}
\usepackage{afterpage}
\usepackage[small]{caption}

\input{tcilatex}
\begin{document}

\title{The adaptivity refines approximate solutions of ill-posed problems
due to the relaxation property}
\author{Larisa Beilina$^{\ast }$ and Michael V. Klibanov$^{\ast \ast }$ \and
$^{\ast }$ Department of Mathematical Sciences \and Chalmers University of
Technology and Gothenburg University \and SE-42196, Gothenburg, Sweden. \and
$^{\ast \ast }$Department of Mathematics and Statistics \and University of
North Carolina at\ Charlotte, \and Charlotte, NC 28223, USA. \and E-mails:
larisa@chalmers.se, mklibanv@uncc.edu}
\maketitle

\begin{abstract}
Adaptive Finite Element Method (adaptivity) is known to be an effective
numerical tool for some ill-posed problems. The key advantage of the
adaptivity is the image improvement with local mesh refinements. A rigorous
proof of this property is the central part of this paper. In terms of
Coefficient Inverse Problems with single measurement data, the authors
consider the adaptivity as the second stage of a two-stage numerical
procedure.\ The first stage delivers a good approximation of the exact
coefficient without an advanced knowledge of a small neighborhood of that
coefficient. This is a necessary element for the adaptivity to start
iterations from. Numerical results for the two-stage procedure are presented
for both computationally simulated and experimental data.
\end{abstract}

\graphicspath{{FIGURES/}
{/chalmers/users/larisa/PAPERS/EXPERIMENT_IP/images/}
 {pics/}}

\textbf{AMS Subject Classification}: 35L10, 35K10, 94A40

\textbf{Key Words: }Adaptive Finite Element Method, relaxation property,
Ill-Posed problems, Coefficient Inverse Problem, numerical studies

\section{Introduction}

\label{sec:1}

For the first time, the relaxation property for the Adaptive Finite Element
Method (adaptivity) for a class of non-linear ill-posed problems was proved
analytically in \cite{BKK}. The relaxation property ensures that the
adaptivity is worth to work with. In short, the relaxation is a rigorously
derived estimate, which shows that the solution computed on a finer mesh is
more accurate than the one computed on a coarser mesh. Unlike classical
Well-Posed problems, this property is not automatic for Ill-Posed problems:
because of the instability of the inversion in the latter case. The main
results of the current paper is Theorem 5.2 (section 5), where a proof,
simpler than the one of \cite{BKK}, is presented. Prior to \cite{BKK} the
relaxation was observed numerically, rather than analytically, in a number
of publications about Coefficient Inverse Problem (CIPs), see, e.g. \cite%
{AB,BJ1,Beil,BJ2,BC,Beilina1,BK2,BK3,BK4}.

In most theorems of this paper (although not in all of them) we consider
only the nonlinear finite dimensional case. The infinitely dimensional case
would likely result in imposing the well known source representation
condition, i.e. assuming that the solution belongs to the range of a certain
compact operator. The latter cannot be effectively verified. On the other
hand, since we are focused on applications of our theory to CIPs, then the
work in a finite dimensional space is well justified by the fact that we
actually work with finite elements, the number of which cannot be too large
in any practical computation.

In our analytical derivations throughout the paper we assume that the noise
level $\delta $ is sufficiently small. This is both a common and natural
assumption in the theory of Ill-Posed problems, especially in the nonlinear
case. Indeed, in principle one can hope to get an accurate solution only if
the noise level is small. However, if the noise is large, then only a very
special procedure, which is designed for a specific problem of interest,
might or might not deliver an accurate solution. Those procedures cannot be
described in the framework of Functional Analysis, since each such procedure
highly depends on many specifics of a problem of interest. On the other
hand, such a procedure took place for the second numerical example of this
paper, which is for experimental data. And noise was quite large in this
case: see comments in the beginning of Section 8.2. This confirms a commonly
known observation that the theory is usually more pessimistic than numerical
examples.

This paper summarizes recent results of the authors on the relaxation
property for the adaptivity for Ill-Posed problems, see \cite{BK1,BKK,KBB}.\
First, results are formulated in the Functional Analysis setting. Next, they
are applied to a CIP for a hyperbolic PDE. Both formulations and proofs of
almost all theorems are modified here, compared with above publications.
%Since journals are sometimes better available for the readers than books, it
%makes sense to have a journal publication of this result.
It is shown in
section 8 (Remark 8.2) that the relaxation property helps to work out the
stopping \ criterion for mesh refinements. Theorems 5.3 and 5.4 as well as
the numerical example of\ Test 1 were not published before.

The essence of the adaptivity consists in the minimization of the Tikhonov
functional on a sequence of locally refined meshes. It is important that due
to local rather than global mesh refinements, the total number of finite
elements is rather moderate. If this number would be very large, then the
corresponding space of finite elements would effectively behave as an
infinitely dimensional one. However, in the case of a moderate number of
finite elements, this space effectively behaves as a finite dimensional one.
Since all norms in finite dimensional spaces are equivalent, then we use the
same norm in the Tikhonov regularization term as the one in the original
space (except of Section 2.1). This is obviously more convenient for both
analysis and numerical studies than the standard case of a stronger norm in
this term \cite{BKok,BK1,Kab,T1,T2}. Numerical results of the current and
previous publications confirm the validity of this approach. Note that
although the finite dimensional version of the original ill-posed problem
might be well posed, at least formally, in the actuality it inherits the
ill-posedness at certain extent. Thus, the use of the regularization term is
still important for the stabilization.

Recall that a minimizer of the Tikhonov functional, if it exists, is called
\emph{regularized solution }of the corresponding equation \cite%
{BKok,BK1,EHN,Kab,T1,T2}. It is well known, however, that Tikhonov
functionals for nonlinear Ill-Posed problems, such as, e.g. CIPs, suffer
from the phenomenon of multiple local minima and ravines. Hence, many
regularized solutions might exist. In addition, there is no guarantee that a
gradient-like or a Newton-like method of minimizing such a functional would
converge to the exact solution $x^{\ast }$, unless the first guess $x_{0}$
would not be sufficiently close to $x^{\ast }.$ In other words, those are
locally convergent methods, so as the adaptivity is. Therefore, the
assumption in some theorems of this paper that the norm $\left\Vert
x_{0}-x^{\ast }\right\Vert $ is sufficiently small is a natural one, and the
goal of the adaptivity is to refine $x_{0}$.

Assuming that the norm $\left\Vert x_{0}-x^{\ast }\right\Vert $ is
sufficiently small, we estimate below the distance between a regularized
solution and the one obtained after adaptive mesh refinements. Next, we
estimate the distance between the latter solution and $x^{\ast }.$ These are
the so-called \textquotedblleft \emph{a posteriori} error estimates"
(Theorems 5.2, 5.3, 6.4 and 6.5 below). This is a new element here.\ Indeed,
in the past publications about the adaptivity for ill-posed problems, a
posteriori error estimates were obtained only for either the Tikhonov
functional or the Lagrangian, rather than for solutions themselves, see,
e.g. \cite{AB,BR,BJ1,Beil,BC,Beilina1,BK2,BK3}.

It follows from the above discussion that, prior to applying the adaptivity
to a CIP, it is necessary to figure out at least one point in a small
neighborhood of the correct solution. Hence, we have developed a two stage
numerical procedure for some CIPs for a hyperbolic PDE. On the first stage,
the so-called \textquotedblleft approximately globally convergent method"
\cite{BK1,BK0,BK5,KFBPS,Klib4,KBK,KBKSNF} delivers the key ingredient of any
locally convergent method: a good approximation $x_{0}$ for the exact
solution $x^{\ast }$. On the second stage, the adaptivity uses this
approximation as a starting point for a refinement \cite{BK1,BK2,BK3,BK4}.

The adaptivity for an ill-posed problem, specifically for a CIP for a
hyperbolic PDE, was first proposed in 2001 in \cite{BJ1}. Also, in
2001 a similar idea was proposed in \cite{BR}, although an example of
a CIP was not considered in \cite{BR}. In both these first
publications the so-called \textquotedblleft Galerkin orthogonality
principle" was used quite essentially. The adaptivity was developed
further in a number of publications, where it was applied to CIPs
\cite{BJ,BJ1,Beil,BJ2,BC,Beilina1}. A posteriori error estimates in an
approximately globally convergent method was derived and an adaptive
globally convergent method was developed at the first time in
\cite{AB}.  In \cite{BKoshev} a posteriori error estimates was
presented and an adaptive finite element method was applied for the
solution of a Fredholm integral equation of a first kind.
We also refer to \cite{Feng}
where the adaptivity was applied to a parameter identification
problem. In a CIP an unknown coefficient of a PDE should be
reconstructed using boundary measurements. In a parameter
identification problem an unknown coefficient is reconstructed
assuming that the solution of the corresponding PDE is given either
everywhere inside of the domain of interest or on a grid inside of
this domain. In the recent publication \cite{Li} the adaptivity was
applied, for the first time, to the classical Cauchy problem for the
Laplace equation and quite accurate images were obtained. Unlike other
works on this topic, both lower and upper error estimates were
obtained in \cite{Li}.

In the sections 2-5 we use the apparatus of the Functional Analysis to
address above items 1-4 for rather general ill-posed problems. In section 6
we deduce from sections 2-5 some results for a CIP for a hyperbolic PDE. In
section 7 we present mesh refinement recommendations. In section 8 we
present numerical results, including ones for real experimental data. In
numerical studies of this paper we use the above mentioned two-stage
numerical procedure.

\section{Minimizing Sequence and a Regularized Solution Versus the First
Guess}

In this section we estimate the distances between terms of the minimizing
sequence of the Tikhonov functional and the exact solution via the distance
between the first guess and the exact solution. In the finite dimensional
case the minimizing sequence is replaced with the regularized solution.

\label{sec:2}

\subsection{The infinitely dimensional case}

\label{sec:2.1}

Let $B,B_{1},B_{2}$ be three Banach spaces. We denote norms in these spaces
respectively as $\left\Vert \cdot \right\Vert ,\left\Vert \cdot \right\Vert
_{1},\left\Vert \cdot \right\Vert _{2}.$ As it is conventional in the theory
of Ill-Posed problems, we assume that $B_{1}\subseteq B,\left\Vert
x\right\Vert \leq C\left\Vert x\right\Vert _{1},\forall x\in B_{1}$ and $%
\overline{B}_{1}=B,$ $C=const.>0$, and the closure $\overline{B}_{1}$ is in
the norm $\left\Vert \cdot \right\Vert .$ Furthermore, we assume that any
bounded set in $B_{1}$ is a compact set in $B$. Let $G\subseteq B_{1}$ be a
set and $\overline{G}$ be its closure in the norm $\left\Vert \cdot
\right\Vert .$ Let $F:\overline{G}\rightarrow B_{2}$ be a one-to-one
operator, which is continuous in terms of norms $\left\Vert \cdot
\right\Vert ,\left\Vert \cdot \right\Vert _{2}.$ Consider the equation
\begin{equation}
F\left( x\right) =y,x\in G.  \label{2.1}
\end{equation}%
As it is usually done in the regularization theory \cite%
{BKok,BK1,EHN,Kab,T1,T2}, we assume that the right hand side of equation (%
\ref{2.1}) is given with a small error $\delta \in \left( 0,1\right) $. We
also assume that there exists an \textquotedblleft ideal" exact solution $%
x^{\ast }$ of (\ref{2.1}) with the \textquotedblleft ideal" exact data $%
y^{\ast }$ (in principle, there might be several exact solutions). Thus, we
assume that
\begin{equation}
F\left( x^{\ast }\right) =y^{\ast },x^{\ast }\in G,\left\Vert y-y^{\ast
}\right\Vert _{2}\leq \delta .  \label{2.2}
\end{equation}

Let $x_{0}\in B_{1}$ be a first guess for the exact solution $x^{\ast }$.
Usually one assumes that $x_{0}$ is located in a small neighborhood of $%
x^{\ast }$. Consider the Tikhonov functional
\begin{equation}
M_{\alpha }\left( x\right) =\frac{1}{2}\left\Vert F\left( x\right)
-y\right\Vert _{2}^{2}+\frac{\alpha }{2}\left\Vert x-x_{0}\right\Vert
_{1}^{2},x,x_{0}\in G,  \label{2.3}
\end{equation}%
where $\alpha \in \left( 0,1\right) $ is the regularization parameter. We
impose a rather conventional assumption that
\begin{equation}
\alpha =\alpha \left( \delta \right) =\delta ^{2\mu },\mu =const.\in \left(
0,1/2\right) .  \label{2.31}
\end{equation}%
The second term in the right hand side of (\ref{2.3}) is called
\textquotedblleft the Tikhonov regularization term". Let
\begin{equation}
m_{\alpha }=\inf_{G}M_{\alpha }\left( x\right) .  \label{2.4}
\end{equation}%
Hence, there exists a minimizing sequence $\left\{ x_{n}^{\alpha }\right\}
_{n=1}^{\infty }\subset G$ such that $\lim_{n\rightarrow \infty }M_{\alpha
}\left( x_{n}^{\alpha }\right) =m_{\alpha }.$ By (\ref{2.2}), (\ref{2.3})
and (\ref{2.4})
\begin{equation}
m_{\alpha }\leq M_{\alpha }\left( x^{\ast }\right) <\delta ^{2}+\alpha
\left\Vert x_{0}-x^{\ast }\right\Vert _{1}^{2}.  \label{2.63}
\end{equation}%
Hence, there exists an integer $N=N\left( \delta ,F\right) \geq 1$ such that
$M_{\alpha }\left( x_{n}^{\alpha \left( \delta \right) }\right) <\delta
^{2}+\alpha \left\Vert x_{0}-x^{\ast }\right\Vert _{1},\forall n\geq N.$
Hence, by (\ref{2.2})
\begin{equation}
\left\Vert x_{n}^{\alpha \left( \delta \right) }\right\Vert _{1}\leq \sqrt{2}%
\left( \delta ^{2\left( 1-\mu \right) }+\left\Vert x_{0}-x^{\ast
}\right\Vert _{1}^{2}\right) ^{1/2}+\left\Vert x_{0}\right\Vert _{1},\forall
n\geq N\left( \delta ,F\right) .  \label{2.7}
\end{equation}%
Suppose that an \emph{a priori} upper estimate of the distance between the
first guess and the exact solution is given,%
\begin{equation}
\left\Vert x_{0}-x^{\ast }\right\Vert _{1}\leq A,A=const.>0,  \label{2.71}
\end{equation}%
where the number $A$ is given. Then (\ref{2.7}) implies that $\left\Vert
x_{n}^{\alpha \left( \delta \right) }\right\Vert _{1}\leq \sqrt{2}\left(
A+1\right) +\left\Vert x_{0}\right\Vert _{1}.$ Consider the set $P\left(
x_{0},A\right) $ defined as
\begin{equation}
P\left( x_{0},A\right) =\left\{ x\in G:\left\Vert x\right\Vert _{1}\leq
\sqrt{2}\left( A+1\right) +\left\Vert x_{0}\right\Vert _{1}\right\} .
\label{2.72}
\end{equation}%
Let $\overline{P}:=\overline{P}\left( x_{0},A\right) $ be its closure in
terms of the norm $\left\Vert \cdot \right\Vert .$ Hence, $\overline{P}%
\subseteq \overline{G}.$ Since the set $P\left( x_{0},A\right) $ is bounded
in terms of the norm $\left\Vert \cdot \right\Vert _{1},$ then $\overline{P}$
is a closed compact set in the space $B$. Consider the range $F\left(
\overline{P}\right) \subset B_{2}$ of the operator $F$ on the set $\overline{%
P}.$ Since the operator $F:\overline{G}\rightarrow B_{2}$ is continuous in
terms of norms $\left\Vert \cdot \right\Vert ,\left\Vert \cdot \right\Vert
_{2}$, then $F\left( \overline{P}\right) $ is a closed compact set in $%
B_{2}. $ Furthermore, since $F$ is one-to-one, then by the foundational
theorem of Tikhonov \cite{BK1,Kab,T1,T2} the inverse operator $%
F^{-1}:F\left( \overline{P}\right) \rightarrow \overline{P}$ is continuous.
Therefore, there exists the modulus of the continuity of the operator $%
F^{-1} $ on the set $F\left( \overline{P}\right) .$ This means that there
exists a function $\omega _{F}\left( z\right) ,z\in \left( 0,\infty \right) $
such that
\begin{eqnarray}
\omega _{F}\left( z\right) &\geq &0,\omega _{F}\left( z_{1}\right) \leq
\omega _{F}\left( z_{2}\right) \text{ if }z_{1}\leq z_{2},\lim_{z\rightarrow
0^{+}}\omega _{F}\left( z\right) =0,  \label{2.8} \\
\left\Vert x_{1}-x_{2}\right\Vert &\leq &\omega _{F}\left( \left\Vert
F\left( x_{1}\right) -F\left( x_{2}\right) \right\Vert _{2}\right) ,\forall
x_{1},x_{2}\in \overline{P}.  \label{2.9}
\end{eqnarray}

Theorem 2.1 compares the distance $\left\Vert x_{0}-x^{\ast }\right\Vert
_{1} $ with the distance between terms of the minimizing sequence and the
exact solution $x^{\ast }$.

\textbf{Theorem 2.1 }(rate of convergence). \emph{Let }$B,B_{1},B_{2}$ \emph{%
be Banach spaces, }$G\subset B_{1}$\emph{\ be a convex open set and }$F:%
\overline{G}\rightarrow B_{2}$\emph{\ be a one-to-one continuous operator in
terms of norms }$\left\Vert \cdot \right\Vert ,\left\Vert \cdot \right\Vert
_{2}.$ \emph{Let conditions (\ref{2.2}), (\ref{2.31}), (\ref{2.4}) and (\ref%
{2.71}) be in place. Then for any number }$\delta \in \left( 0,1\right) $
\emph{there exists an integer }$N=N\left( \delta ,F\right) \geq 1$\emph{\
such that }%
\begin{equation}
\left\Vert x_{n}^{\alpha \left( \delta \right) }-x^{\ast }\right\Vert \leq
\omega _{F}\left( 2\delta ^{\mu }\sqrt{A+1}\right) ,\forall n\geq N\left(
\delta ,F\right) .  \label{2.91}
\end{equation}

\textbf{Proof}. Using (\ref{2.2}) and (\ref{2.63}), we obtain for $n\geq
N\left( \delta ,F\right) $%
\begin{eqnarray}
\left\Vert F\left( x_{n}^{\alpha \left( \delta \right) }\right) -F\left(
x^{\ast }\right) \right\Vert _{2} &=&\left\Vert F\left( x_{n}^{\alpha \left(
\delta \right) }\right) -y+y-F\left( x^{\ast }\right) \right\Vert _{2}
\notag \\
&\leq &\left\Vert F\left( x_{n}^{\alpha \left( \delta \right) }\right)
-y\right\Vert _{2}+\left\Vert y-y^{\ast }\right\Vert _{2}\leq \left[
2M_{\alpha }\left( x_{n}^{\alpha \left( \delta \right) }\right) \right]
^{1/2}+\delta  \label{2.92} \\
&\leq &\left( \delta ^{2}+\delta ^{2\mu }\left\Vert x_{0}-x^{\ast
}\right\Vert _{1}^{2}\right) ^{1/2}+\delta \leq 2\delta ^{\mu }\left(
1+\left\Vert x_{0}-x^{\ast }\right\Vert _{1}^{2}\right) ^{1/2}\leq 2\delta
^{\mu }\sqrt{A+1}.  \notag
\end{eqnarray}%
By (\ref{2.63}), (\ref{2.71}) and (\ref{2.72}) $x^{\ast }\in \overline{P}.$
Therefore, (\ref{2.9}) and (\ref{2.92}) imply (\ref{2.91}). $\square $

Theorem 2.1 estimates the distance $\left\Vert x_{n}^{\alpha \left( \delta
\right) }-x^{\ast }\right\Vert $ via the distance $\left\Vert x_{0}-x^{\ast
}\right\Vert _{1}$ between the first guess and the exact solution for any $%
\delta \in \left( 0,1\right) .$ Still, it is natural to ensure that the
distance between terms of the minimizing sequence and the exact solution is
strictly less than the distance between the first guess $x_{0}$ and the
exact solution. This can be ensured only for sufficiently small values of
the noise level $\delta .$Although Corollary 2.1 has a similarity with the
well known convergence theorem of the minimizing sequence for the Tikhonov
functional (see, e.g. page 33 in \cite{BK1}), still in that theorem only a
subsequence converges rather than the entire sequence. Besides, estimate (%
\ref{2.10}) is useful by its own right, and also the convergence rate (\ref%
{2.91}), from which (\ref{2.10}) is derived, seems to be new.

\textbf{Corollary 2.1}. \emph{Let }$B,B_{1},B_{2}$ \emph{be Banach spaces, }$%
G\subset B_{1}$\emph{\ be a convex open set and }$F:\overline{G}\rightarrow
B_{2}$\emph{\ be a one-to-one continuous operator in terms of norms }$%
\left\Vert \cdot \right\Vert ,\left\Vert \cdot \right\Vert _{2}.$ \emph{Let
conditions (\ref{2.2}), (\ref{2.31}), (\ref{2.4}) and (\ref{2.71}) be in
place. Let }$\xi \in \left( 0,1\right) $ \emph{be an arbitrary number.
Assume first that }$x_{0}\neq x^{\ast }.$\emph{\ Then there exists a
sufficiently small number }$\delta _{0}=$\emph{\ }$\delta _{0}\left( F,A,\mu
,\xi \right) \in \left( 0,1\right) $\emph{\ such that }%
\begin{equation}
\left\Vert x_{n}^{\alpha \left( \delta \right) }-x^{\ast }\right\Vert \leq
\xi \left\Vert x_{0}-x^{\ast }\right\Vert ,\forall \delta \in \left(
0,\delta _{0}\right) ,n\geq N\left( \delta ,F\right) .  \label{2.10}
\end{equation}%
\emph{In the case }$x_{0}=x^{\ast }$\emph{\ (\ref{2.10}) should be replaced
with}
\begin{equation}
\left\Vert x_{n}^{\alpha \left( \delta \right) }-x^{\ast }\right\Vert \leq
\xi ,\forall \delta \in \left( 0,\delta _{0}\right) ,n\geq N\left( \delta
,F\right) .  \label{2.11}
\end{equation}%
\emph{In particular, if }$\delta =0,$\emph{\ then }$\delta _{0}$\emph{\
should be replaced with a sufficiently small number }$\alpha _{0}\in \left(
0,1\right) $\emph{\ and \textquotedblleft }$\delta \in \left( 0,\delta
_{0}\right) "$\emph{\ should be replaced with }$\alpha \in \left( 0,\alpha
_{0}\right) .$

\textbf{Proof}. First, let $x_{0}\neq x^{\ast }.$ By (\ref{2.8}) there
exists a sufficiently small number $\delta _{0}\left( F,A,\mu ,\xi \right)
\in \left( 0,1\right) $ such that $\omega _{F}\left( 2\delta ^{\mu }\sqrt{A+1%
}\right) \leq \xi \left\Vert x_{0}-x^{\ast }\right\Vert ,\forall \delta \in
\left( 0,\delta _{0}\right) .$ Combining this with (\ref{2.91}), we obtain (%
\ref{2.10}).

Let now $x_{0}=x^{\ast }.$ Then again there exists a sufficiently small
number $\delta _{0}\left( F,A,\mu ,\xi \right) \in \left( 0,1\right) $ such
that $\omega _{F}\left( 2\delta ^{\mu }\sqrt{A+1}\right) \leq \xi .$
Combining this with (\ref{2.91}), we obtain (\ref{2.11}). $\square $

\subsection{The finite dimensional case}

\label{sec:2.2}

Consider now the finite dimensional real valued Hilbert space.\ Compared
with subsection 2.1, the main new point here is that the minimizing sequence
is replaced with a minimizer, which exists. This case is of our main
interest in the current paper because standard piecewise linear finite
elements form a finite dimensional space. Unlike the above, we now use the
same norm in the regularization term as in the original space. This is
because all norms are equivalent in a finite dimensional space.
Nevertheless, since the finite dimensional version of the original ill-posed
problem \textquotedblleft inherits" the ill-posedness, at certain extent, it
is still important to use the regularization term for the stabilization.

Let $H$ and $H_{2}$ be two real valued Hilbert spaces and $\dim H<\infty .$
Norms and scalar products in these spaces denote respectively as $\left\Vert
\cdot \right\Vert ,\left( ,\right) ,\left\Vert \cdot \right\Vert _{2},\left(
,\right) _{2}.$ Let $G\subset H$ be an open bounded set and $F:\overline{G}%
\rightarrow H_{2}$ be a continuous operator. We again consider equations (%
\ref{2.1}), (\ref{2.2}), where $x^{\ast }\in G,y,y^{\ast }\in H_{2}$. The
functional $M_{\alpha }\left( x\right) $ in (\ref{2.3}) is now replaced with
the functional $J_{\alpha }\left( x\right) ,$
\begin{equation}
J_{\alpha }\left( x\right) =\frac{1}{2}\left\Vert F\left( x\right)
-y\right\Vert _{2}^{2}+\frac{\alpha }{2}\left\Vert x-x_{0}\right\Vert
^{2},x\in \overline{G},x_{0}\in G.  \label{2.14}
\end{equation}

\emph{\ }The following lemma follows immediately from Weierstrass theorem.

\textbf{Lemma 2.1}. \emph{Let }$F$\emph{\ be the operator defined above in
this section. Then} \emph{there exists a regularized solution }$x_{\alpha
}\in \overline{G}$,
\begin{equation}
\inf_{\overline{G}}J_{\alpha }\left( x\right) =\min_{\overline{G}}J_{\alpha
}\left( x\right) =J_{\alpha }\left( x_{\alpha }\right) .  \label{2.15}
\end{equation}

Although a similar result is valid for the case when the set $G$ is
unbounded, we do not formulate it here since we do not need it. The
following theorem follows immediately from Theorem 2.1 and Corollary 2.1.

\textbf{Theorem 2.2}. \emph{Let Hilbert spaces }$H,H_{2}$\emph{, the set }$%
G\subset H$\emph{\ and the operator }$F:\overline{G}\rightarrow H_{2}$\emph{%
\ be a one-to-one continuous operator. Let conditions (\ref{2.2}), (\ref%
{2.71}), (\ref{2.14}) and (\ref{2.15}) be in place. Then for any number }$%
\delta \in \left( 0,1\right) $%
\begin{equation*}
\left\Vert x_{\alpha \left( \delta \right) }-x^{\ast }\right\Vert \leq
\omega _{F}\left( 2\delta ^{\mu }\sqrt{A+1}\right) .
\end{equation*}%
\emph{\ Let }$\xi \in \left( 0,1\right) $\emph{\ be an arbitrary constant.
Then there exists a sufficiently small number }$\delta _{0}=$\emph{\ }$%
\delta _{0}\left( F,A,\mu ,\xi \right) \in \left( 0,1\right) $\emph{\ such
that for all }$\delta \in \left( 0,\delta _{0}\right) $\emph{\ }%
\begin{equation*}
\left\Vert x_{\alpha \left( \delta \right) }-x^{\ast }\right\Vert \leq
\left\{
\begin{array}{c}
\xi \left\Vert x_{0}-x^{\ast }\right\Vert ,\text{ if }x_{0}\neq x^{\ast },
\\
\xi ,\text{ if }x_{0}=x^{\ast }.%
\end{array}%
\right.
\end{equation*}%
\emph{\ In particular, if }$\delta =0,$\emph{\ then }$\delta _{0}$\emph{\
should be replaced with a sufficiently small number }$\alpha _{0}\in \left(
0,1\right) $\emph{\ and \textquotedblleft }$\delta \in \left( 0,\delta
_{0}\right) "$\emph{\ should be replaced with }$\alpha \in \left( 0,\alpha
_{0}\right) .$

\section{The Local Strong Convexity of the Tikhonov Functional (\protect\ref%
{2.14})}

\label{sec:3}

In \cite{Ramlau} the local strong convexity of the Tikhonov functional was
established for the case when the underlying operator $F$\ has the second
continuous Fr\'{e}chet derivative and the source representation condition is
in place. In this section we prove the local strong convexity of the
Tikhonov functional (\ref{2.14}) for the case when the operator $F$ has the
first continuous Fr\'{e}chet derivative and the source representation
condition is not imposed.

Let $H$ and $H_{2}$ be two real valued Hilbert spaces. Let scalar products
and norms in them be respectively $\left( ,\right) ,\left\Vert \cdot
\right\Vert $ and $\left( ,\right) _{2},\left\Vert \cdot \right\Vert _{2}.$
Let $\mathcal{L}\left( H,H_{2}\right) $ be the the space of all bounded
linear operators mapping $H$ into $H_{2}$ and let $\left\Vert \cdot
\right\Vert _{\mathcal{L}}$ be the norm in $\mathcal{L}\left( H,H_{2}\right)
.$ Although we do not assume here that $H$ is finite dimensional, we still
use the same norm $\left\Vert x-x_{0}\right\Vert $ in the regularization
term in (\ref{2.14}) as the one in the original space $H$, rather than a
stronger norm as in (\ref{2.3}). This is again because our true goal is to
work in a finite dimensional space of finite elements in the adaptivity
(section 1). For any $a>0$ and for any $x\in H$ denote $V_{a}\left( x\right)
=\left\{ z\in H:\left\Vert x-z\right\Vert <a\right\} .$ First, we formulate
the following well known theorem.

\textbf{Theorem 3.1.} \cite{Minoux}. \emph{Let }$G\subseteq H$\emph{\ be a
convex open set and }$L:G\rightarrow \mathbb{R}$\emph{\ be a functional.
Suppose that this functional has the Fr\'{e}chet derivative }$L^{\prime
}\left( x\right) \in \mathcal{L}\left( H,\mathbb{R}\right) $\emph{\ for
every point }$x\in G.$\emph{\ Then the strong convexity of }$L$\emph{\ on
the set }$G$\emph{\ with the strong convexity constant }$\rho >0$ \emph{is
equivalent with the following condition \ }
\begin{equation}
\left( L^{\prime }\left( x\right) -L^{\prime }\left( z\right) ,x-z\right)
\geq 2\rho \left\Vert x-z\right\Vert ^{2},\forall x,z\in G.  \label{3.1}
\end{equation}%
\emph{\ }

\textbf{Theorem 3.2. }\emph{Let} $G\subseteq H$\emph{\ be a convex open set
and }$F:\overline{G}\rightarrow H_{2}$\emph{\ be an operator. Let }$x^{\ast
}\in G$\emph{\ be an exact solution of equation (\ref{2.1}) with the exact
data }$y^{\ast }$. \emph{Let }$V_{1}\left( x^{\ast }\right) \subset G$\emph{%
\ and let (\ref{2.2}) holds. Assume that for every }$x\in V_{1}\left(
x^{\ast }\right) $\emph{\ the operator }$F$\emph{\ has the Fr\'{e}chet
derivative }$F^{\prime }\left( x\right) \in \mathcal{L}\left( H,H_{2}\right)
.$\emph{\ Suppose that this derivative is uniformly bounded and Lipschitz
continuous in }$V_{1}\left( x^{\ast }\right) $\emph{, i.e. }
\begin{eqnarray}
\left\Vert F^{\prime }\left( x\right) \right\Vert _{\mathcal{L}} &\leq
&N_{1},\text{ }\forall x\in V_{1}\left( x^{\ast }\right) ,  \label{3.2} \\
\left\Vert F^{\prime }\left( x\right) -F^{\prime }\left( z\right)
\right\Vert _{\mathcal{L}} &\leq &N_{2}\left\Vert x-z\right\Vert ,\text{ }%
\forall x,z\in V_{1}\left( x^{\ast }\right) ,  \label{3.3}
\end{eqnarray}%
\emph{\ } \emph{where }$N_{1},N_{2}=const.>0.$\emph{\ Let }
\begin{eqnarray}
\alpha &=&\alpha \left( \delta \right) =\delta ^{2\mu },\quad \forall \delta
\in \left( 0,1\right) ,  \label{3.4} \\
\mu &=&const.\in \left( 0,\frac{1}{4}\right) .  \label{3.5}
\end{eqnarray}%
\emph{\ } \emph{Then there exists a} \emph{sufficiently small number }$%
\delta _{0}=\delta _{0}\left( N_{1},N_{2},\mu \right) \in \left( 0,1\right) $%
\emph{\ such that for all }$\delta \in \left( 0,\delta _{0}\right) $\emph{\
the functional }$J_{\alpha \left( \delta \right) }\left( x\right) $\emph{\
is strongly convex in the neighborhood }$V_{\delta ^{3\mu }}\left( x^{\ast
}\right) $\emph{\ of }$x^{\ast }$\emph{\ with the strong convexity constant }%
$\alpha /4.$ \emph{In the noiseless case with }$\delta =0$\emph{\ one should
replace \textquotedblleft }$\delta _{0}=\delta _{0}\left( N_{1},N_{2},\mu
\right) \in \left( 0,1\right) "$\emph{\ with }$\alpha _{0}=\alpha _{0}\left(
N_{1},N_{2}\right) \in \left( 0,1\right) $\emph{\ to be sufficiently small
and require that }$\alpha \in \left( 0,\alpha _{0}\right) .$

We refer to \cite{BK1,BKK} for the proof of Theorem 3.2 since it is space
consuming. Consider now the finite dimensional case.

\textbf{Theorem 3.3}. \emph{Let }$\dim H<\infty ,G\subset H$\emph{\ be an
open bounded convex set, and the rest of conditions of Theorem 3.2 holds.
Let in (\ref{2.14}) the first guess }$x_{0}$ \emph{\ for the exact solution }%
$x^{\ast }$\emph{\ be so accurate that }
\begin{equation}
\left\Vert x_{0}-x^{\ast }\right\Vert <\frac{\delta ^{3\mu }}{3}.
\label{3.14}
\end{equation}%
\emph{Then there exists a sufficiently small number }$\delta _{0}=\delta
_{0}\left( N_{1},N_{2},\mu \right) \in \left( 0,1\right) $\emph{\ such that
for every }$\delta \in \left( 0,\delta _{0}\right) $\emph{\ and for }$\alpha
=\alpha \left( \delta \right) $\emph{\ satisfying (\ref{3.4}) there exists
unique regularized solution }$x_{\alpha \left( \delta \right) }$\emph{\ of
equation (\ref{2.1}) on the set }$G.$\emph{\ Furthermore, }$x_{\alpha \left(
\delta \right) }\in V_{\delta ^{3\mu }}\left( x^{\ast }\right) .$\emph{\ In
addition, the gradient method of the minimization of the functional }$%
J_{\alpha \left( \delta \right) }\left( x\right) ,$\emph{\ which starts at }$%
x_{0},$\emph{\ converges to }$x_{\alpha \left( \delta \right) }.$ \emph{%
Also, if the operator }$F$\emph{\ is one-to-one on }$V_{1}\left( x^{\ast
}\right) $\emph{, then }$x_{\alpha \left( \delta \right) }\in V_{\delta
^{3\mu }/3}\left( x^{\ast }\right) .$ \emph{In the noiseless case with }$%
\delta =0$\emph{\ one should replace \textquotedblleft }$\delta _{0}=\delta
_{0}\left( N_{1},N_{2},\mu \right) \in \left( 0,1\right) "$\emph{\ with }$%
\alpha _{0}=\alpha _{0}\left( N_{1},N_{2}\right) \in \left( 0,1\right) $%
\emph{\ to be sufficiently small and require that }$\alpha \in \left(
0,\alpha _{0}\right) .$

\textbf{Proof}. By Lemma 2.1 there exists a minimizer $x_{\alpha \left(
\delta \right) }\in \overline{G}$ of the functional $J_{\alpha \left( \delta
\right) }$. We have $J_{\alpha \left( \delta \right) }\left( x_{\alpha
\left( \delta \right) }\right) \leq J_{\alpha \left( \delta \right) }\left(
x^{\ast }\right) .$ Also, $\left\Vert x_{\alpha \left( \delta \right)
}-x_{0}\right\Vert \geq \left\Vert x_{\alpha \left( \delta \right) }-x^{\ast
}\right\Vert -\left\Vert x_{0}-x^{\ast }\right\Vert .$ Hence, using (\ref%
{2.2}), (\ref{2.14}) and (\ref{3.14}), we\emph{\ }obtain that there exists a
sufficiently small number $\delta _{0}=\delta _{0}\left( N_{1},N_{2},\mu
\right) \in \left( 0,1\right) $ such that for every $\delta \in \left(
0,\delta _{0}\right) $%
\begin{equation*}
\left\Vert x_{\alpha \left( \delta \right) }-x^{\ast }\right\Vert \leq \frac{%
\delta }{\sqrt{\alpha }}+2\left\Vert x^{\ast }-x_{0}\right\Vert <\delta
^{1-\mu }+\frac{2}{3}\delta ^{3\mu }=\frac{2}{3}\delta ^{3\mu }\left( 1+%
\frac{3}{2}\delta ^{1-4\mu }\right) <\frac{2}{3}\delta ^{3\mu }\cdot \frac{3%
}{2}=\delta ^{3\mu }.
\end{equation*}%
Hence, $x_{\alpha \left( \delta \right) }\in V_{\delta ^{3\mu }}\left(
x^{\ast }\right) .$ Since by Theorem 3.2 the functional $J_{\alpha }$ is
strongly convex on the set $V_{\delta ^{3\mu }}\left( x^{\ast }\right) $ and
the minimizer $x_{\alpha \left( \delta \right) }\in V_{\delta ^{3\mu
}}\left( x^{\ast }\right) ,$ then this minimizer is unique. Furthermore,
since by (\ref{3.14}) the point $x_{0}\in V_{\delta ^{3\mu }}\left( x^{\ast
}\right) ,$ then it is well known that the gradient method with its starting
point at $x_{0}$ converges to $x_{\alpha \left( \delta \right) }$.

Let now the operator $F$ be one-to-one. Let $\xi \in \left( 0,1\right) $ be
an arbitrary number and $x_{0}\neq x^{\ast }$. By Theorem 2.2 we can choose
a smaller number $\delta _{0}=\delta _{0}\left( N_{1},N_{2},\mu ,\xi \right)
$ such that
\begin{equation*}
\left\Vert x_{\alpha \left( \delta \right) }-x^{\ast }\right\Vert \leq \xi
\left\Vert x_{0}-x^{\ast }\right\Vert ,\forall \delta \in \left( 0,\delta
_{0}\right) .
\end{equation*}%
Hence, (\ref{3.14}) implies that $x_{\alpha \left( \delta \right) }\in
V_{\delta ^{3\mu }/3}\left( x^{\ast }\right) .$ If $x_{0}=x^{\ast },$ then
by Theorem 2.2 $\left\Vert x_{\alpha \left( \delta \right) }-x^{\ast
}\right\Vert \leq \xi .$ Choosing $\xi \in \left( 0,\delta ^{3\mu }/3\right)
,$ we again obtain that $x_{\alpha \left( \delta \right) }\in V_{\delta
^{3\mu }/3}\left( x^{\ast }\right) .$ The noiseless case is similar. $%
\square $

\section{The Space of Finite Elements}

\label{sec:4}

To prove the relaxation property of the adaptivity, we need to introduce the
space of finite elements.
 Let $\Omega
\subset \mathbb{R}^{n},n=2,3$ be a bounded domain.
Consider a discretization of  $\Omega$ by an unstructured mesh $T$ using
non-overlapping tetrahedral elements in $ \mathbb{R}^{3}$ and
triangles in $\mathbb{R}^{2}$ such that $T=K_1, ...,K_l$, where $l$ is
the number of elements in $\Omega$, and
\begin{equation*}
D = \cup_{K \in T} K=K_1 \cup K_2...\cup K_l.
\end{equation*}
We obtain a polygonal domain $D$
and assume for brevity that $D=\Omega.$
We associate with the triangulation $T$ the mesh function $h=h(x)$
which is a piecewise-constant function such that
\begin{equation*}
h(x)= h_K ~~~ \forall K \in T,
\end{equation*}
where $h_K$ is the diameter of $K$ which we define as the longest
side of $K$.
 Following section 76.4 of \cite{ERJ}, consider piecewise linear
 functions $\left\{ e_{j}\left( x,T\right) \right\} _{j=1}^{N}\subset
 C\left( \overline{\Omega }\right) $, which are called \emph{test
   functions}. Functions $\left\{ e_{j}\left( x,T\right) \right\}_{j=1}^{N}$ are linearly independent in $\Omega $. Here, $N$ is
   the global number of nodes in the mesh $T$. Let $\left\{
   N_{i}\right\} $ be the set of nodal points of
   triangle/tetrahedra  $K$ for all $K \in T$. Then
\begin{equation*}
e_{j}\left( N_{i}, T\right) =\left\{
\begin{array}{c}
1,i=j, \\
0,i\neq j.%
\end{array}%
\right.
\end{equation*}
We introduce the finite element space
$V_h$ as
\begin{equation}
V_h = \big\{ v(x) \in V: v \in C(\Omega),~ v|_{K} \in P_1(K)~\forall K \in T \big\},
\end{equation}
where $P_1(K)$ denotes the set of piecewise-linear functions on $K$ with
\begin{equation*}
V= \big\{v(x): v(x) \in L_2(\Omega) \big\}.
\end{equation*}
The finite dimensional finite element space $V_h$ is constructed such
that $V_h \subset V$.

Let $r$
be the radius of the maximal circle/sphere inscribed in $K$. We impose
the shape regularity assumption for all triangles/tetrahedra uniformly for
all possible triangulations $T$ which we consider. Specifically, we assume
that
\begin{equation}
a_{1}\leqslant h_K \leqslant r a_{2},\quad
a_{1},a_{2}=const.>0,\text{ }\forall K \in T,\text{ }\forall ~T,
\label{4.1}
\end{equation}%
where numbers $a_{1},a_{2}$ are independent on the triangulation $T$. Let $%
h_{\max }\left( T\right) $ and $h_{\min }\left( T\right) $ be respectively
the maximal and minimal diameters of triangles/tetrahedra of the
triangulation $T$. We assume evrywhere below that
\begin{equation}
\frac{h_{\min }\left( T\right) }{h_{\max }\left( T\right) }\leq  c_T,\forall T
\label{4.10}
\end{equation}
for a certain positive constant $c_T$. Obviously, the number of all possible
triangulations satisfying (\ref{4.1}), (\ref{4.10}) is finite. Thus, we
introduce the following finite dimensional linear space $H,$
\begin{equation*}
H=\bigcup\limits_{T}V_h\left( T\right) ,\text{ }\forall T\text{ \ satisfying (%
\ref{4.1}), (\ref{4.10}).}
\end{equation*}%
Hence,
\begin{equation}
\dim H<\infty ,\text{ }H\subset \left( C\left( \overline{\Omega }\right)
\cap H^{1}\left( \Omega \right) \right) ,\text{ }\partial _{x_{i}}f\in
L_{\infty }\left( \Omega \right) ,\text{ }\forall f\in H.  \label{4.2}
\end{equation}%
In (\ref{4.2}) "$\subset $" means the inclusion of sets. We equip $H$ with
the same inner product as the one in $L_{2}\left( \Omega \right) .$ Denote $%
\left( ,\right) $ and $\left\Vert \cdot \right\Vert $ the inner product and
the norm in $H$ respectively, $\left\Vert f\right\Vert _{H}:=\left\Vert
f\right\Vert _{L_{2}\left( \Omega \right) }:=\left\Vert f\right\Vert ,$ $
\forall f\in H.$ Everywhere below $H$ is this space. We view the space $H$
as an \textquotedblleft ideal\textquotedblright\ space of very fine finite
elements, which cannot be reached in practical computations. At the same
time, all other spaces of finite elements we work with below are subspaces
of $H.$ In particular, this means that we assume without further mentioning
that (\ref{4.1}) and (\ref{4.10}) are valid for all meshes considered below.

Keeping in mind the mesh refinement process in the adaptivity, we now
explain how do we construct triangulations $\left\{ T_{n}\right\} $ as well
as corresponding subspaces $\left\{ M_{n}\right\} $ of the space $H$ which
correspond to mesh refinements. Consider the first triangulation $T_{1}$
with rather coarse mesh. We set $M_{1}:=V_h\left( T_{1}\right) \subset H.$
Suppose that the pair $\left( T_{n},M_{n}\right) $ is constructed after $n$
mesh refinements and that the basis functions in the space $M_{n}$ are $%
\left\{ e_{j}\left( x,T_{n}\right) \right\} _{j=1}^{N_n}.$
We now want to refine the mesh again. We define the pair $\left(
T_{n+1},M_{n+1}\right) $ as follows. We refine the mesh in the standard
manner as it is usually done when working with triangular/tetrahedron finite
elements. When doing so, we keep (\ref{4.1}). Hence, we obtain both the
triangulation $T_{n+1}$ and the corresponding test functions $\left\{
e_{j}\left( x,T_{n+1}\right) \right\} _{j=1}^{N_{n+1}}$. It
is well known that test functions $\left\{ e_{j}\left( x,T_{n}\right)
\right\} _{j=1}^{N_n}$ are linearly dependent from new
test functions $\left\{ e_{j}\left( x,T_{n+1}\right) \right\}
_{j=1}^{N_{n+1}}.$ Thus, we define the subspace $M_{n+1}$ as
\begin{equation*}
M_{n+1}:=\func{Span}\left( \left\{ e_{j}\left( x,T_{n+1}\right) \right\}
_{j=1}^{N_{n+1}}\right) .
\end{equation*}%
Therefore, we have obtained a finite set of linear subspaces $\left\{
M_{n}\right\} _{n=1}^{N}$ of the space $H.$ Each subspace $M_{n}$
corresponds to the mesh refinement number $n,M_{n+1}\diagdown M_{n}\neq
\varnothing $ and
\begin{equation*}
M_{n}\subset M_{n+1}\subset H,n\in \left[ 1,N-1\right] .
\end{equation*}

Let $I$  be the identity operator on $H$. For any subspace $M\subset
H,$ let $P_{M}:H\rightarrow M$ be the orthogonal projection operator
of the space $H$ onto its subspace $M$. Denote for brevity
$P_{n}:=P_{M_{n}}.$ Let $h_{n}$ be the maximal grid step size of
$T_{n}$. Hence, $h_{n+1}\leqslant h_{n}.$ Let $f_{n}^{I}$ be the
standard interpolant of the function $f\in H$ on triangles/tetrahedra
of $T_{n},$ see section 76.4 of \cite{ERJ}. It can be easily derived
from formula (76.3) of \cite{ERJ} that
\begin{equation}
\text{ }\left\Vert f-f_{n}^{I}\right\Vert \leq K\left\Vert \nabla
f\right\Vert _{L_{\infty }\left( \Omega \right) }h_{n},\forall f\in H,
\label{4.3}
\end{equation}%
where $K=K\left( \Omega ,r,a_{1},a_{2}\right) =const.>0.$ Since $%
f_{n}^{I}\in H,\forall f\in H,$ then by one of well known properties of
orthogonal projection operators,
\begin{equation}
\left\Vert f-P_{n}f\right\Vert \leqslant \left\Vert f-f_{n}^{I}\right\Vert ,%
\text{ }\forall f\in H.  \label{4.4}
\end{equation}%
Hence, (\ref{4.3}) and (\ref{4.4}) imply that with a different constant $%
K=K\left( \Omega ,r,a_{1},a_{2}\right) >0$
\begin{equation}
\left\Vert f-P_{n}f\right\Vert \leqslant K\left\Vert \nabla f\right\Vert
_{L_{\infty }\left( \Omega \right) }h_{n},\forall f\in H.  \label{4.5}
\end{equation}%
Since $H$ is a finite dimensional space in which all norms are equivalent,
it is convenient for us to rewrite (\ref{4.5}) with a different constant $%
K=K\left( \Omega ,r,,a_{1},a_{2}\right) >0$ as
\begin{equation}
\left\Vert x-P_{n}x\right\Vert \leqslant K\left\Vert x\right\Vert h_{n},%
\text{ }\forall x\in H.  \label{4.6}
\end{equation}

\section{Relaxation}

\label{sec:5}

Since we sequentially minimize the Tikhonov functional on subspaces $\left\{
M_{n}\right\} _{n=1}^{N}$ in the adaptivity procedure, then we need to
establish first the existence of a minimizer on each of these subspaces. In
this section the set $G$ and the operator $F$ are the same as in Theorem
3.3, and the functional $J_{\alpha }\left( x\right) $ is the same as in (\ref%
{2.14}). Theorem 5.1 ensures both existence and uniqueness of the minimizer
of the functional $J_{\alpha }$ on each subspace of the space $H,$ as long
as the maximal grid step size of finite elements, which are involved in that
subspace, is sufficiently small.

\textbf{Theorem 5.1. }\emph{Let conditions of Theorem 3.3 hold. In
particular, let the operator }$F:\overline{G}\rightarrow H_{2}$ \emph{be
one-to-one.} \emph{Let }$M\subseteq H$\emph{\ be a subspace of }$H$\emph{\
and let }$V_{\delta ^{3\mu }}\left( x^{\ast }\right) \cap M\neq \varnothing $%
.\emph{\ Assume that }$\left\Vert x^{\ast }\right\Vert \leq B$, \emph{where
the number }$B>0$\emph{\ is known in advance. Suppose that the maximal grid
step size }$\widetilde{h}$\emph{\ of finite elements of }$M$\emph{\ be so
small that }
\begin{equation}
\widetilde{h}\leq \frac{\delta ^{4\mu }}{5BN_{2}K},  \label{5.5}
\end{equation}%
\emph{where }$K$\emph{\ is the constant in (\ref{4.6}). Furthermore, assume
that the first guess }$x_{0}$\emph{\ for the exact solution }$x^{\ast }$
\emph{in the functional }$J_{\alpha \left( \delta \right) }$ \emph{is so
accurate that (\ref{3.14}) is in place. Then} \emph{there exists a} \emph{%
sufficiently small number }$\delta _{0}=\delta _{0}\left( N_{1},N_{2},\mu
\right) \in \left( 0,1\right) $\emph{\ such that for every }$\delta \in
\left( 0,\delta _{0}\right) $ \emph{there exists unique minimizer }$%
x_{M,\alpha \left( \delta \right) }\in G\cap M$\emph{\ of the functional }$%
J_{\alpha }$\emph{\ on the set }$G\cap M.$\ \emph{Furthremore, }$x_{M,\alpha
\left( \delta \right) }\in V_{\delta ^{3\mu }}\left( x^{\ast }\right) \cap
M. $\emph{\ In addition, the functional }$J_{\alpha }\left( x\right) $\emph{%
\ is strongly convex on the set }$V_{\delta ^{3\mu }}\left( x^{\ast }\right)
\cap M$ \emph{with the strong convexity constant }$\alpha \left( \delta
\right) /4.$\emph{\ Let }$x_{\alpha \left( \delta \right) }\in V_{\delta
^{3\mu }/3}\left( x^{\ast }\right) $ \emph{be the regularized solution of
equation (\ref{2.1}), which is guaranteed by Theorem 3.3.\ Then} \emph{the
following a posteriori error estimate holds}
\begin{equation*}
\left\Vert x_{M,\alpha \left( \delta \right) }-x_{\alpha \left( \delta
\right) }\right\Vert \leq \frac{2}{\delta ^{2\mu }}\left\Vert J_{\alpha
}^{\prime }\left( x_{M,\alpha \left( \delta \right) }\right) \right\Vert .
\end{equation*}

Note that since in Theorem 5.1 $V_{1}\left( x^{\ast }\right) \subset G$ and $%
V_{\delta ^{3\mu }}\left( x^{\ast }\right) \cap M\neq \varnothing $, then $%
G\cap M\neq \varnothing .$ We do not prove this theorem here and refer
instead to Theorem 4.9.2 of \cite{BK1}; also see Theorem 3.2 of \cite{BKK}
for a similar result.

\textbf{Theorem 5.2 }(relaxation)\textbf{.} \emph{Let }$M_{n}\subset H$\emph{%
\ be the subspace obtained after }$n$\emph{\ mesh refinements, as described
in section 4. Let }$h_{n}$\emph{\ be the maximal grid step size of the
subspace }$M_{n}$\emph{.\ Suppose that all conditions of Theorem 5.1 hold
with the only exception that the subspace }$M$\emph{\ is replaced with }$%
M_{n}$\emph{\ and the inequality (\ref{5.5}) is replaced with }
\begin{equation}
h_{n}\leq \frac{\delta ^{4\mu }}{5BN_{2}K}.  \label{5.7}
\end{equation}%
\emph{Let }$\delta \in \left( 0,\delta _{0}\right) ,$\emph{\ where the
number }$\delta _{0}\in \left( 0,1\right) $\emph{\ is defined in Theorem
5.1. Also, let }$V_{\delta ^{3\mu }}\left( x^{\ast }\right) \cap M_{1}\neq
\varnothing .$\emph{\ Let }$x_{n}\in V_{\delta ^{3\mu }}\left( x^{\ast
}\right) \cap M_{n}$\emph{\ be the unique minimizer of the functional }$%
J_{\alpha }\left( x\right) $\emph{\ in (\ref{2.14}) on the set }$G\cap M_{n}$%
\emph{\ (Theorem 5.1). Let }$x_{\alpha \left( \delta \right) }\in V_{\delta
^{3\mu }/3}\left( x^{\ast }\right) $\emph{\ be the unique regularized
solution (Theorem 3.3). Assume that }
\begin{equation}
x_{n}\neq x_{\alpha \left( \delta \right) },  \label{5.8}
\end{equation}%
\emph{\ i.e. }$x_{\alpha \left( \delta \right) }\notin M_{n\text{ }},$\emph{%
\ meaning that the regularized solution is not yet reached after }$n$\emph{\
mesh refinements. Let }$\eta \in \left( 0,1\right) $\emph{. Then one can
choose the maximal grid size }$h_{n+1}=h_{n+1}\left( N_{1},N_{2},\delta
,B,K,\eta \right) \in \left( 0,h_{n}\right] $\emph{\ of the mesh refinement
number }$\left( n+1\right) $\emph{\ so small that }
\begin{equation}
\left\Vert x_{n+1}-x_{\alpha \left( \delta \right) }\right\Vert \leq \eta
\left\Vert x_{n}-x_{\alpha \left( \delta \right) }\right\Vert ,  \label{5.9}
\end{equation}%
\emph{where }$x_{n+1}\in V_{\delta ^{3\mu }}\left( x^{\ast }\right) \cap
M_{n+1}$\emph{\ is the unique minimizer of the functional (\ref{2.14}) on
the set }$G\cap M_{n+1}$\emph{. Hence,}
\begin{equation}
\left\Vert x_{n+1}-x_{\alpha \left( \delta \right) }\right\Vert \leq \eta
^{n}\left\Vert x_{1}-x_{\alpha \left( \delta \right) }\right\Vert .
\label{5.10}
\end{equation}

\textbf{Proof.} In this proof we denote for brevity\textbf{\ }$\alpha \left(
\delta \right) :=\alpha .$ Since $V_{\delta ^{3\mu }}\left( x^{\ast }\right)
\cap M_{1}\neq \varnothing ,M_{1}\subseteq M_{n}$ and $V_{\delta ^{3\mu
}}\left( x^{\ast }\right) \subset V_{1}\left( x^{\ast }\right) \subset G,$
then $\left( V_{\delta ^{3\mu }}\left( x^{\ast }\right) \cap M_{n}\right)
\subset \left( V_{1}\left( x^{\ast }\right) \cap M_{n+1}\right) \neq
\varnothing .$ Since by Theorem 3.2 the functional (\ref{2.14}) is strongly
convex on the set $V_{\delta ^{3\mu }}\left( x^{\ast }\right) $ with the
strong convexity constant $\alpha /4,$ then Theorem 3.1 implies that
\begin{equation}
\frac{\alpha }{2}\left\Vert x_{n+1}-x_{\alpha \left( \delta \right)
}\right\Vert ^{2}\leq \left( J_{\alpha }^{\prime }\left( x_{n+1}\right)
-J_{\alpha }^{\prime }\left( x_{\alpha \left( \delta \right) }\right)
,x_{n+1}-x_{\alpha \left( \delta \right) }\right) .  \label{5.11}
\end{equation}%
Since $x_{n+1}$ is the minimizer on $G\cap M_{n+1}$ and $x_{\alpha }$ is the
minimizer on the set $G,$ then
\begin{equation}
\left( J_{\alpha }^{\prime }\left( x_{n+1}\right) ,z\right) =0,\text{ }%
\forall z\in M_{n+1};J_{\alpha }^{\prime }\left( x_{\alpha \left( \delta
\right) }\right) =0.  \label{5.12}
\end{equation}%
Relations (\ref{5.12}) justify the application of the Galerkin orthogonality
principle \cite{BR,BJ1}. By (\ref{5.12})
\begin{equation}
\left( J_{\alpha }^{\prime }\left( x_{n+1}\right) -J_{\alpha }^{\prime
}\left( x_{\alpha \left( \delta \right) }\right) ,x_{n+1}-P_{n+1}x_{\alpha
\left( \delta \right) }\right) =0.  \label{5.13}
\end{equation}%
Next, $x_{n+1}-x_{\alpha \left( \delta \right) }=\left(
x_{n+1}-P_{n+1}x_{\alpha \left( \delta \right) }\right) +\left(
P_{n+1}x_{\alpha \left( \delta \right) }-x_{\alpha \left( \delta \right)
}\right) .$ Hence, (\ref{5.11}) and (\ref{5.13}) imply that
\begin{equation}
\frac{\alpha }{2}\left\Vert x_{n+1}-x_{\alpha \left( \delta \right)
}\right\Vert ^{2}\leq \left( J_{\alpha }^{\prime }\left( x_{n+1}\right)
-J_{\alpha }^{\prime }\left( x_{\alpha \left( \delta \right) }\right)
,P_{n+1}x_{\alpha \left( \delta \right) }-x_{\alpha \left( \delta \right)
}\right) .  \label{5.14}
\end{equation}%
It follows from (\ref{3.3}) that conditions (\ref{5.7}) and (\ref{5.8})
imply that
\begin{equation}
\left\Vert J_{\alpha }^{\prime }\left( x_{n+1}\right) -J_{\alpha }^{\prime
}\left( x_{\alpha \left( \delta \right) }\right) \right\Vert \leq
N_{3}\left\Vert x_{n+1}-x_{\alpha \left( \delta \right) }\right\Vert
\label{5.15}
\end{equation}%
with a constant $N_{3}=N_{3}\left( N_{1},N_{2}\right) >0.$ Also, by (\ref%
{4.6})
\begin{equation}
\left\Vert x_{\alpha \left( \delta \right) }-P_{n+1}x_{\alpha \left( \delta
\right) }\right\Vert \leq K\left\Vert x_{\alpha \left( \delta \right)
}\right\Vert h_{n+1}.  \label{5.16}
\end{equation}%
Using the Cauchy-Schwarz inequality as well as (\ref{3.4}), (\ref{5.15}) and
(\ref{5.16}), we obtain from (\ref{5.14})
\begin{equation}
\left\Vert x_{n+1}-x_{\alpha \left( \delta \right) }\right\Vert \leq \frac{%
2KN_{3}}{\delta ^{2\mu }}\left\Vert x_{\alpha \left( \delta \right)
}\right\Vert h_{n+1}.  \label{5.17}
\end{equation}

Since by one of conditions of Theorem 5.1 we have an \emph{a priori} known
upper estimate $\left\Vert x^{\ast }\right\Vert \leq B,$ we now can estimate
the norm $\left\Vert x_{\alpha \left( \delta \right) }\right\Vert $. Since
by Theorem 3.3 $x_{\alpha \left( \delta \right) }\in V_{\delta ^{3\mu
}/3}\left( x^{\ast }\right) ,$ then
\begin{equation*}
\left\Vert x_{\alpha \left( \delta \right) }\right\Vert \leq \left\Vert
x_{\alpha \left( \delta \right) }-x^{\ast }\right\Vert +\left\Vert x^{\ast
}\right\Vert \leq \frac{\delta ^{3\mu }}{3}+B.
\end{equation*}%
Hence, (\ref{5.17}) becomes
\begin{equation}
\left\Vert x_{n+1}-x_{\alpha \left( \delta \right) }\right\Vert \leq \frac{%
2KN_{3}}{\delta ^{2\mu }}\left( \frac{\delta ^{3\mu }}{3}+B\right) h_{n+1}.
\label{5.18}
\end{equation}

Let $\eta _{n}\in \left( 0,1\right) $ be an arbitrary number. Since $%
\left\Vert x_{n}-x_{\alpha \left( \delta \right) }\right\Vert \neq 0,$ then
we can choose $h_{n+1}=h_{n+1}\left( N_{2},\delta ,A,K\right) \in \left(
0,h_{n}\right] $ so small that
\begin{equation}
\frac{2KN_{3}}{\delta ^{2\mu }}\left( \frac{\delta ^{3\mu }}{3}+B\right)
h_{n+1}\leq \eta \left\Vert x_{n}-x_{\alpha \left( \delta \right)
}\right\Vert .  \label{5.19}
\end{equation}%
Comparing (\ref{5.19}) with (\ref{5.18}), we obtain the target estimate (\ref%
{5.9}). $\square $

Theorem 5.2 provides an estimate of the distance between points $x_{n+1}$
obtained via adaptive mesh refinement and the regularized solution. We now
estimate how far are these points from the exact solution $x^{\ast }.$%
Theorem 5.3 follows immediately from Theorem 2.2 and (\ref{5.10}).

\textbf{Theorem 5.3}. \emph{Let conditions of Theorem 5.2 hold. Let }$\delta
\in \left( 0,\delta _{0}\right) ,$\emph{\ where the number }$\delta _{0}\in
\left( 0,1\right) $\emph{\ is defined in Theorem 5.1. Then there exists a
decreasing sequence of maximal grid step sizes }$\left\{ h_{k}\right\}
_{k=1}^{n+1}$\emph{\ such that }%
\begin{equation}
\left\Vert x_{k+1}-x^{\ast }\right\Vert \leq \eta ^{k}\left\Vert
x_{1}-x_{\alpha \left( \delta \right) }\right\Vert +\omega _{F}\left(
2\delta ^{\mu }\sqrt{A+1}\right) ,k=1,...,n,  \label{5.201}
\end{equation}%
\emph{where the number }$A$\emph{\ is defined in (\ref{2.71}) and the
function }$\omega _{F}$\emph{\ is defined in (\ref{2.8}), (\ref{2.9}). In
particular, let }$\xi \in \left( 0,1\right) $\emph{\ be an arbitrary number.
Then there exists a sufficiently small number }$\delta _{1}=\delta
_{1}\left( N_{1},N_{2},\mu ,\xi \right) \in \left( 0,\delta _{0}\right] $%
\emph{\ and a decreasing sequence of maximal grid step sizes }$\left\{
h_{k}\right\} _{k=1}^{n+1}$\emph{\ such that for all }$\delta \in \left(
0,\delta _{1}\right) $ and for $k\in \left[ 1,n\right] $\emph{\ }%
\begin{equation}
\left\Vert x_{k+1}-x^{\ast }\right\Vert \leq \eta ^{k}\left\Vert
x_{1}-x_{\alpha \left( \delta \right) }\right\Vert +\left\{
\begin{array}{c}
\xi \left\Vert x_{0}-x^{\ast }\right\Vert ,\text{ if }x_{0}\neq x^{\ast },
\\
\xi ,\text{ if }x_{0}=x^{\ast }.%
\end{array}%
\right.  \label{5.20}
\end{equation}

Since $h_{n+1}$ is the maximal grid step size in the entire domain $\Omega ,$
it seems to be at the first glance that Theorems 5.2, 5.3 are about mesh
refinements in the entire domain $\Omega $ rather than about local mesh
refinements in subdomains, as it is the case in the adaptivity. Assuming
that conditions of Theorem 5.2 hold, we now show that local mesh refinements
are also covered by this theorem. Suppose that the domain $\Omega $ is split
in two subdomains, $\Omega =\Omega _{1}\cup \Omega _{2},\Omega _{1}\cap
\Omega _{2}=\varnothing .$ Assume that the function $x_{0}$ is changing
slowly in $\Omega _{1}$ and has some \textquotedblleft bumps" in $\Omega
_{2}.$ These bumps correspond to small inclusions.\ It is these inclusions
rather than slowly changing functions, which are of the main applied
interest in imaging. Indeed, those small abnormalities model, e.g. land
mines, tumors, etc. Hence, it is reasonable to assume that $x^{\ast }$ is
also changing slowly in $\Omega _{1}.$ Next, because of Theorem 2.2 and
because all norms in $H$ are equivalent, it is reasonable to assume that the
regularized solution $x_{\alpha }$ is also changing slowly in $\Omega _{1}.$%
Thus, inequality (\ref{5.22}) of Theorem 5.4 is a reasonable one.
Furthermore, it is reasonable to assume that mesh refinements do not take
place in $\Omega _{1}$, but only in $\Omega _{2}.$

\textbf{Theorem 5.4} (relaxation for local mesh refinements). \emph{Assume
that conditions of Theorem 5.2 hold. Let }$h^{\left( 1\right) }$\emph{\ be
the maximal grid step size in }$\Omega _{1}.$\emph{Then there exists a
sufficiently small number }$\delta _{0}=\delta _{0}\left( N_{1},N_{2},\mu
\right) \in \left( 0,1\right) $\emph{\ and a decreasing sequence of maximal
grid step sizes }$\left\{ \widetilde{h}_{k}\right\} _{k=1}^{n+1}$\emph{\
such that if the norm }$\left\Vert \nabla x_{\alpha \left( \delta \right)
}\right\Vert _{L_{\infty }\left( \Omega _{1}\right) }$\emph{\ is so small
that with the constant }$N_{3}=N_{3}\left( N_{1},N_{2}\right) >0$ \emph{from
(\ref{5.15}) }%
\begin{equation}
\frac{2KN_{3}}{\delta ^{2\mu }}\left\Vert \nabla x_{\alpha \left( \delta
\right) }\right\Vert _{L_{\infty }\left( \Omega _{1}\right) }h^{\left(
1\right) }\leq \frac{\eta }{2}\left\Vert x_{k}-x_{\alpha \left( \delta
\right) }\right\Vert ,k=1,...,n,  \label{5.22}
\end{equation}%
\emph{then (\ref{5.201}) and (\ref{5.20}) hold with the replacement of }$%
\left\{ h_{k}\right\} _{k=1}^{n+1}$\emph{\ with }$\left\{ \widetilde{h}%
_{k}\right\} _{k=1}^{n+1}.$

\textbf{Proof}. By (\ref{5.14}) and (\ref{5.15})
\begin{eqnarray}
\left\Vert x_{k+1}-x_{\alpha \left( \delta \right) }\right\Vert &\leq &\frac{%
2N_{3}}{\delta ^{2\mu }}\left\Vert x_{\alpha \left( \delta \right)
}-P_{k+1}x_{\alpha \left( \delta \right) }\right\Vert =  \label{5.23} \\
&&\frac{2N_{3}}{\delta ^{2\mu }}\left( \left\Vert x_{\alpha \left( \delta
\right) }-P_{k+1}x_{\alpha \left( \delta \right) }\right\Vert _{L_{2}\left(
\Omega _{1}\right) }+\left\Vert x_{\alpha \left( \delta \right)
}-P_{k+1}x_{\alpha \left( \delta \right) }\right\Vert _{L_{2}\left( \Omega
_{2}\right) }\right) .  \notag
\end{eqnarray}%
By (\ref{4.5}) and (\ref{5.22})
\begin{equation}
\frac{2N_{3}}{\delta ^{2\mu }}\left\Vert x_{\alpha \left( \delta \right)
}-P_{k+1}x_{\alpha \left( \delta \right) }\right\Vert _{L_{2}\left( \Omega
_{1}\right) }\leq \frac{2KN_{3}}{\delta ^{2\mu }}\left\Vert \nabla x_{\alpha
\left( \delta \right) }\right\Vert _{L_{\infty }\left( \Omega _{1}\right)
}h^{\left( 1\right) }\leq \frac{\eta }{2}\left\Vert x_{k}-x_{\alpha \left(
\delta \right) }\right\Vert .  \label{5.24}
\end{equation}%
Next, we obtain similarly with (\ref{5.19})
\begin{equation}
\frac{2N_{3}}{\delta ^{2\mu }}\left\Vert x_{\alpha \left( \delta \right)
}-P_{k+1}x_{\alpha \left( \delta \right) }\right\Vert _{L_{2}\left( \Omega
_{2}\right) }\leq \frac{\eta }{2}\left\Vert x_{k}-x_{\alpha \left( \delta
\right) }\right\Vert .  \label{5.25}
\end{equation}%
It follows from (\ref{5.23})-(\ref{5.25}) that
\begin{equation*}
\left\Vert x_{k+1}-x_{\alpha \left( \delta \right) }\right\Vert \leq \eta
\left\Vert x_{k}-x_{\alpha \left( \delta \right) }\right\Vert ,k=1,...,n.
\end{equation*}%
Hence, (\ref{5.10}) holds. Finally, (\ref{5.20}) follows from (\ref{5.10})
and Theorem 2.2. $\square $

\textbf{Remark 5.1}. Theorems 5.3 and 5.4 claim that the accuracy of the
solution improves with mesh refinements, i.e., the relaxation takes place.
Comparison of (\ref{5.8}) with (\ref{5.201}) and (\ref{5.20}) shows that the
solution is adaptively refined until reaching the regularized solution $%
x_{\alpha \left( \delta \right) }.$ It is important that by Theorem 2.2 the
accuracy of $x_{\alpha \left( \delta \right) }$ is better than the accuracy
of the first guess $x_{0}$. Indeed, this ensures that it is worthy to work
with the adaptivity in order to improve the accuracy of the regularized
solution via mesh refinements.

\section{Adaptivity for a Coefficient Inverse Problem}

\label{sec:6}

We now reformulate some of above theorems for the case of a specific CIP. To
save space, we do not prove theorems of this section. Instead, we point to
those results of Chapter 4 of \cite{BK1} from which these theorems can be
easily derived.

\subsection{Coefficient Inverse Problem and Tikhonov functional}

\label{sec:6.1}

Let $\Omega \subset \mathbb{R}^{3}$ be a convex bounded domain with the
boundary $\partial \Omega \in C^{3}.$ Let the point $x_{0}\notin \overline{%
\Omega }.$ For $T>0$ denote $Q_{T}=\Omega \times \left( 0,T\right)
,S_{T}=\partial \Omega \times \left( 0,T\right) .$ Let $d>1$ be a certain
number, $\omega \in \left( 0,1\right) $ be a sufficiently small number, and
the function $c\left( x\right) \in C\left( \mathbb{R}^{3}\right) $ be such
that
\begin{equation}
c\left( x\right) \in \left( 1-\omega ,d+\omega \right) \text{ in }\Omega
,c\left( x\right) =1\text{ outside of }\Omega .  \label{6.1}
\end{equation}%
Below we specify $c\left( x\right) $ more. Consider the solution $u\left(
x,t\right) $ of the following Cauchy problem%
\begin{eqnarray}
c\left( x\right) u_{tt} &=&\Delta u,x\in \mathbb{R}^{3},t\in \left(
0,T\right) ,  \label{6.2} \\
u\left( x,0\right) &=&0,u_{t}\left( x,0\right) =\delta \left( x-x_{0}\right)
.  \label{6.3}
\end{eqnarray}%
Equation (\ref{6.2}) governs propagation of acoustic waves, in which case $%
c\left( x\right) =1/b^{2}\left( x\right) ,$ where $b\left( x\right) $ is the
sound speed and $u\left( x,t\right) $ is the amplitude of the acoustic wave
\cite{Colton}. In addition, (\ref{6.2}) governs propagation of the
electromagnetic field in 2-d, in which case $c\left( x\right) =\varepsilon
_{r}\left( x\right) $ is the spatially distributed dielectric constant and $%
u\left( x,t\right) $ is one of components of the electric field \cite{RMC}.
Although in the latter application equation (\ref{6.2}) is valid only in
2-d, we have successfully used this equation to work with experimental data,
which are obviously in 3-d, see \cite{BK1,BK4,KFBPS} and section 8. This was
explained in Test 4 of \cite{Beilina2}. It was shown in this test that the
component of the electric field, which was initially sent in a rather simple
medium, dominates two other components. It was also shown that the
propagation of the dominated component is well governed by equation (\ref%
{6.2}).

\textbf{Remark 6.1.} An alternative to the point source in (\ref{6.3}) is
the incident plane wave in the case when it is initialized at the plane $%
\left\{ x_{3}=x_{3,0}\right\} $ such that $\left\{ x_{3}=x_{3,0}\right\}
\cap \overline{\Omega }=\varnothing .$ The formalism of derivations below is
similar in this case. In our derivations below we focus on (\ref{6.3}),
because this is the most convenient case for derivations. However, in
numerical studies we use the incident plane wave, because this case has
shown a better performance than the point source.

\textbf{Coefficient Inverse Problem (}CIP\textbf{)}. Let conditions (\ref%
{6.1})-(\ref{6.3}) hold. Assume that the coefficient $c\left( x\right) $ is
unknown inside the domain $\Omega $. Determine this coefficient for $x\in
\Omega ,$ assuming that the following function $g\left( x,t\right) $ is known%
\begin{equation}
u\mid _{S_{T}}=g\left( x,t\right) .  \label{6.4}
\end{equation}

The function $g\left( x,t\right) $ can be interpreted as the result of
measurements of the wave field $u\left( x,t\right) $ at the boundary of the
domain of interest $\Omega .$ Since the function $c\left( x\right) =1$
outside of $\Omega ,$ then (\ref{6.2})-(\ref{6.4}) imply%
\begin{eqnarray*}
u_{tt} &=&\Delta u,\left( x,t\right) \in \left( \mathbb{R}^{3}\diagdown
\Omega \right) \times \left( 0,T\right) , \\
u\left( x,0\right) &=&u_{t}\left( x,0\right) =0,x\in \mathbb{R}^{3}\diagdown
\Omega ,u\mid _{S_{T}}=g\left( x,t\right) .
\end{eqnarray*}%
Solving this initial boundary value problem in the domain $\left\{ \left(
x,t\right) \in \left( \mathbb{R}^{3}\diagdown \Omega \right) \times \left(
0,T\right) \right\} ,$ we uniquely obtain the Neumann boundary condition $%
p\left( x,t\right) $ for the function $u,$%
\begin{equation}
\partial _{n}u\mid _{S_{T}}=p\left( x,t\right) .  \label{6.5}
\end{equation}

CIPs are quite complex problems. Hence, to handle them, one naturally needs
to impose some simplifying assumptions. In this particular CIP our theory of
the adaptivity is not working unless we replace the $\delta -$function in (%
\ref{6.3}) by a smooth function, which approximates $\delta \left(
x-x_{0}\right) $ in the distribution sense. Let $\varkappa \in \left(
0,1\right) $ be a sufficiently small number. We replace $\delta \left(
x-x_{0}\right) $ in (\ref{6.3}) with the function $\delta _{\varkappa
}\left( x-x_{0}\right) ,$%
\begin{equation}
\delta _{\varkappa }\left( x-x_{0}\right) =\left\{
\begin{array}{c}
C_{\varkappa }\exp \left( \frac{1}{\left\vert x-x_{0}\right\vert
^{2}-\varkappa ^{2}}\right) ,\left\vert x-x_{0}\right\vert <\varkappa , \\
0,\left\vert x-x_{0}\right\vert >\varkappa ,%
\end{array}%
\right. \int\limits_{\mathbb{R}^{3}}\delta _{\varkappa }\left(
x-x_{0}\right) dx=1.  \label{6.6}
\end{equation}%
We assume that $\varkappa $ is so small that
\begin{equation}
\delta _{\varkappa }\left( x-x_{0}\right) =0\text{ in }\overline{\Omega }.
\label{6.7}
\end{equation}%
We now introduce state and adjoint problems. Let $\zeta \in \left(
0,1\right) $ be a sufficiently small number. Consider the function $z_{\zeta
}\in C^{\infty }\left[ 0,T\right] $ such that%
\begin{equation}
z_{\zeta }\left( t\right) =\left\{
\begin{array}{c}
1,t\in \left[ 0,T-2\zeta \right] , \\
0,t\in \left[ T-\zeta ,T\right] , \\
\text{ between }0\text{ and }1\text{ for }t\in \left[ 0,T-2\zeta ,T-\zeta %
\right] .%
\end{array}%
\right.  \label{6.8}
\end{equation}

\textbf{State Problem}. Find the solution $v\left( x,t\right) $ of the
following initial boundary value problem
\begin{equation}
\begin{split}
c\left( x\right) v_{tt}-\Delta v& =0\text{ in }Q_{T}, \\
v(x,0)& =v_{t}(x,0)=0, \\
\partial _{n}v\mid _{S_{T}}& =p\left( x,t\right) .
\end{split}
\label{6.9}
\end{equation}

\textbf{Adjoint Problem}. Find the solution $\lambda \left( x,t\right) $ of
the following initial boundary value problem with the reversed time
\begin{equation}
\begin{split}
\ c\left( x\right) \lambda _{tt}-\Delta \lambda & =0\text{ in }Q_{T}, \\
\lambda (x,T)& =\lambda _{t}(x,T)=0, \\
\partial _{n}\lambda \mid _{S_{T}}& =z_{\zeta }\left( t\right) \left(
g-v\right) \left( x,t\right) .
\end{split}
\label{6.10}
\end{equation}

Here functions $v\in H^{1}\left( Q_{T}\right) $ and $\lambda \in H^{1}\left(
Q_{T}\right) $ are weak solutions of problems (\ref{6.9}) and (\ref{6.10})
respectively. In fact, we need a higher smoothness of these functions, which
we specify below. In (\ref{6.9}) and (\ref{6.10}) functions $g$ and $p$ are
the ones from (\ref{6.4}) and (\ref{6.5}) respectively. Hence, to solve the
adjoint problem, one should solve the state problem first. The function $%
z_{\zeta }\left( t\right) $ is introduced to ensure the validity of
compatibility conditions at $\left\{ t=T\right\} $ in (\ref{6.10}). The
Tikhonov functional for the above CIP is%
\begin{equation}
E_{\alpha }(c)=\frac{1}{2}\int\limits_{S_{T}}(v\mid
_{S_{T}}-~g(x,t))^{2}z_{\zeta }\left( t\right) d\sigma dt+\frac{1}{2}\alpha
\int\limits_{\Omega }(c-c_{glob})^{2}dx,  \label{6.11}
\end{equation}%
where the function $c_{glob}\in C\left( \overline{\Omega }\right) $ is the
approximate solution obtained by our approximately globally convergent
numerical method on the first stage of our two stage numerical procedure
(section 1) and $\alpha $ is the small regularization parameter.

State and adjoint problems are concerned only with the domain $\Omega $
rather than with the entire space $\mathbb{R}^{3}.$ We define the space $Z$
as%
\begin{equation*}
Z=\left\{ f:f\in C\left( \overline{\Omega }\right) \cap H^{1}\left( \Omega
\right) ,c_{x_{i}}\in L_{\infty }\left( \Omega \right) ,i=1,2,3\right\}
,\left\Vert f\right\Vert _{Z}=\left\Vert f\right\Vert _{C\left( \overline{%
\Omega }\right) }+\sum\limits_{i=1}^{3}\left\Vert f_{x_{i}}\right\Vert
_{L_{\infty }\left( \Omega \right) }.
\end{equation*}%
Clearly $H\subset Z$ as a set. To apply the theory of above sections, we
express in subsection 6.2 the function $c(x)$ via standard piecewise linear
finite elements. Hence, we assume below that $c\in Y,$ where%
\begin{equation}
Y=\left\{ c\in Z:c\in \left( 1-\omega ,d+\omega \right) \right\} .
\label{6.14}
\end{equation}%
To find the Fr\'{e}chet derivative of the functional $E_{\alpha }(c)$, we
need to find Fr\'{e}chet derivatives of functions solutions $v,\lambda $ of
problems (\ref{6.9}), (\ref{6.10}). This, in turn requires a higher
smoothness of functions $p,g$ \cite{BK1,BK3}. Theorem 6.1 can be easily
derived from a combination of Theorems 4.7.1, 4.7.2 and 4.8 of \cite{BK1} as
well as from Theorems 3.1, 3.2 of \cite{BK3}.

\textbf{Theorem 6.1}. \emph{Let }$\Omega \subset \mathbb{R}^{3}$\emph{\ be a
convex bounded domain with the boundary }$\partial \Omega \in C^{2}$\emph{\
and such that there exists a function }$a\in C^{2}\left( \overline{\Omega }%
\right) $\emph{\ such that }$a\mid _{\partial \Omega }=0,\partial _{n}a\mid
_{\partial \Omega }=1.$\emph{\ Assume that there exist functions }$P\left(
x,t\right) ,\Phi \left( x,t\right) $\emph{\ such that}%
\begin{eqnarray*}
P &\in &H^{6}\left( Q_{T}\right) ,\Phi \in H^{5}\left( Q_{T}\right)
;\partial _{n}P\mid _{S_{T}}=p\left( x,t\right) ,\partial _{n}\Phi \mid
_{S_{T}}=z_{\zeta }\left( t\right) g\left( x,t\right) , \\
\partial _{t}^{j}P\left( x,0\right) &=&\partial _{t}^{j}\Phi \left(
x,0\right) =0,j=1,2,3,4.
\end{eqnarray*}%
\emph{Then for every function }$c\in Y$\emph{\ functions }$v,\lambda \in
H^{2}\left( Q_{T}\right) ,$ \emph{where }$v,\lambda $\emph{\ are solutions
of state and adjoint problems (\ref{6.9}), (\ref{6.10}). Also, for every }$%
c\in Y$ \emph{there exists Fr\'{e}chet derivative }$E_{\alpha }^{\prime }(c)$%
\emph{\ of the Tikhonov functional }$E_{\alpha }:Y\rightarrow \mathbb{R}$%
\emph{\ in (\ref{6.11}) and }%
\begin{equation}
E_{\alpha }^{\prime }(c)\left( x\right) =\alpha \left( c-c_{glob}\right)
\left( x\right) -\int\limits_{0}^{T}\left( u_{t}\lambda _{t}\right) \left(
x,t\right) ~dt:=\alpha \left( c-c_{glob}\right) \left( x\right) +y\left(
x\right) .  \label{6.15}
\end{equation}%
\emph{Functions }$E_{\alpha }^{\prime }(c)\left( x\right) ,y\left( x\right)
\in C\left( \overline{\Omega }\right) $\emph{\ and there exists a constant }$%
D=D\left( \Omega ,a,d,\omega ,z_{\zeta }\right) >0$\emph{\ such that }%
\begin{equation}
\left\Vert y\right\Vert _{C\left( \overline{\Omega }\right) }\leq \left\Vert
c\right\Vert _{C\left( \overline{\Omega }\right) }^{2}\exp \left( DT\right)
\left( \left\Vert P\right\Vert _{H^{6}\left( Q_{T}\right) }^{2}+\left\Vert
\Phi \right\Vert _{H^{5}\left( Q_{T}\right) }^{2}\right) .  \label{6.16}
\end{equation}%
\emph{The functional of the Fr\'{e}chet derivative }$E_{\alpha }^{\prime
}(c) $\emph{\ acts on any function }$b\in Z$\emph{\ as}%
\begin{equation*}
E_{\alpha }^{\prime }(c)\left( b\right) =\int\limits_{\Omega }E_{\alpha
}^{\prime }(c)\left( x\right) b\left( x\right) dx.
\end{equation*}

\subsection{Relaxation property for the functional $E_{\protect\alpha }(c)$}

\label{sec:6.2}

In this section we use Theorems 5.2, 5.4 to derive the relaxation property
for the for the specific functional $E_{\alpha }(c)$ for our CIP. The first
step is to define the operator $F$ for our specific case. Set $G:=Y\cap H$.
We consider the set $G$ as the subset of the space $H$ with the same norm as
the one in $H$. In particular, $\overline{G}=\left\{ c\left( x\right) \in
H:c\left( x\right) \in \left[ 1-\omega ,d+\omega \right] \text{ for }x\in
\overline{\Omega }\right\} .$ Let $H_{2}:=L_{2}\left( S_{T}\right) .$ We
define the operator $F$ as
\begin{equation}
F:\overline{G}\rightarrow H_{2},F\left( c\right) \left( x,t\right) =z_{\zeta
}\left( t\right) \left[ g\left( x,t\right) -v\left( x,t,c\right) \right] ,%
\text{ }\left( x,t\right) \in S_{T},  \label{6.17}
\end{equation}%
where the function $v:=v\left( x,t,c\right) $ is the weak solution (\ref%
{6.13}) of the state problem (\ref{6.9}), $g$ is the function in (\ref{6.4})
and $z_{\zeta }\left( t\right) $ is the function defined in (\ref{6.8}). For
any function $b\in H$ consider the weak solution $\widetilde{u}\left(
x,t,c,b\right) \in H^{1}\left( Q_{T}\right) $ of the following initial
boundary value problem%
\begin{eqnarray*}
c\left( x\right) \widetilde{u}_{tt} &=&\Delta \widetilde{u}-b\left( x\right)
v_{tt},\left( x,t\right) \in Q_{T}, \\
\widetilde{u}\left( x,0\right) &=&\widetilde{u}_{t}\left( x,0\right) =0,%
\widetilde{u}\mid _{S_{T}}=0.
\end{eqnarray*}%
Theorem 6.2 can be easily derived from a combination of Theorems 4.7.2 and
4.10 of \cite{BK1}.

\textbf{Theorem 6.2.} \emph{Let }$\Omega \subset \mathbb{R}^{3}$\emph{\ be a
convex bounded domain with the boundary }$\partial \Omega \in C^{2}.$\emph{\
Suppose that there exist functions }$a\left( x\right) ,P\left( x,t\right)
,\Phi \left( x,t\right) $\emph{\ satisfying conditions of Theorem 6.1. Then
the function }$\widetilde{u}\left( x,t,c,b\right) \in H^{2}\left(
Q_{T}\right) .$\emph{\ Also, the operator }$F$\emph{\ in (\ref{6.17}) has
the Fr\'{e}chet derivative }$F^{\prime }\left( c\right) \left( b\right) ,$\
\begin{equation*}
F^{\prime }\left( c\right) \left( b\right) =-z_{\zeta }\left( t\right)
\widetilde{u}\left( x,t,c,b\right) \mid _{S_{T}},\forall c\in G,\forall b\in
H.
\end{equation*}%
\emph{Let }$B=B\left( \Omega ,a,d,\omega ,z_{\zeta }\right) >0$\emph{\ be
the constant of Theorem 6.1. Then}
\begin{equation*}
\left\Vert F^{\prime }\left( c\right) \right\Vert _{\mathcal{L}}\leq \exp
\left( CT\right) \left( \left\Vert P\right\Vert _{H^{6}\left( Q_{T}\right)
}+\left\Vert \Phi \right\Vert _{H^{5}\left( Q_{T}\right) }\right) ,\text{ }%
\forall c\in G.
\end{equation*}%
\ \emph{In addition, the operator }$F^{\prime }\left( c\right) $\emph{\ is
Lipschitz continuous,}
\begin{equation*}
\left\Vert F^{\prime }\left( c_{1}\right) -F^{\prime }\left( c_{2}\right)
\right\Vert _{\mathcal{L}}\leq \exp \left( CT\right) \left( \left\Vert
P\right\Vert _{H^{6}\left( Q_{T}\right) }+\left\Vert \Phi \right\Vert
_{H^{5}\left( Q_{T}\right) }\right) \left\Vert c_{1}-c_{2}\right\Vert ,\text{
}\forall c_{1},c_{2}\in G.
\end{equation*}%
\emph{\ }

Following (\ref{2.2}), we introduce the error of the level $\delta $ in the
data $g(x,t)$ in (\ref{6.4}). So, we assume that
\begin{equation}
g(x,t)=g^{\ast }(x,t)+g_{\delta }(x,t);\text{ }g^{\ast },g_{\delta }\in
L_{2}\left( S_{T}\right) ,\left\Vert g_{\delta }\right\Vert _{L_{2}\left(
S_{T}\right) }\leq \delta .  \label{6.18}
\end{equation}%
where $g^{\ast }(x,t)$ is the exact data and the function $g_{\delta }(x,t)$
represents the error in these data. To make sure that the operator $F$ is
one-to-one, we need to refer to a uniqueness theorem for our CIP.\ However,
uniqueness results for multidimensional CIPs with single measurement data
are currently known only under the assumption that at least one of initial
conditions does not equal zero in the entire domain $\overline{\Omega },$
which is not our case. All these theorems were proven by the method, which
was originated in 1981 in three papers \cite{BukhKlib,Bukh,Klib1}; also see,
e.g. \cite{Bukh1,Klib2,Klib3,KT,Klib6,Klib4} as well as sections 1.10, 1.11
of the book \cite{BK1} and references cited there for some follow up
publications of those authors about this method. This method is based on
Carleman estimates. Although many other researchers have published about
this method, we do not cite those works here, because the topic of
uniqueness is not a focus of the current paper. We refer to surveys \cite%
{Klib4,Y} for more references. Lifting the above assumption is a long
standing and well known open question, see \cite{Klib6} for a recent partial
answer to this question. Nevertheless, because of applications, it makes
sense to develop numerical methods for the above CIP, regardless on the
absence of proper uniqueness theorems. Therefore, we introduce Assumption
6.1.

\textbf{Assumption 6.1.} \emph{The operator }$F\left( c\right) $\emph{\
defined in (\ref{6.17}) is one-to-one. }

Theorem 6.3 follows from Theorems 3.3, 6.1 and 6.2. Note that if a function $%
c\in H$ is such that $c\in \left[ 1,d\right] ,$ then by (\ref{6.14}) $c\in
G. $

\textbf{Theorem 6.3.} \emph{Let }$\Omega \subset \mathbb{R}^{3}$\emph{\ be a
convex bounded domain with the boundary }$\partial \Omega \in C^{3}.$\emph{\
Suppose that there exist functions }$a\left( x\right) ,P\left( x,t\right)
,\Phi \left( x,t\right) $\emph{\ satisfying conditions of Theorem 6.1. Let
Assumption 6.1 and condition (\ref{6.18}) hold. Let the function }$v=v\left(
x,t,c\right) \in H^{2}\left( Q_{T}\right) $\emph{\ in (\ref{6.11}) be the
solution of the state problem (\ref{6.9}) for the function }$c\in G$\emph{.
Assume that there exists the exact solution }$c^{\ast }\in G,c^{\ast }\left(
x\right) \in \left[ 1,d\right] $\emph{\ of the equation }$F\left( c^{\ast
}\right) =0$\emph{\ for the case when\ in (\ref{6.18}) the function }$g$%
\emph{\ is replaced with the function }$g^{\ast }$\emph{. Let in (\ref{6.18}%
) }%
\begin{equation*}
\alpha =\alpha \left( \delta \right) =\delta ^{2\mu },\mu =const.\in \left(
0,1/4\right) .
\end{equation*}%
\emph{\ Also, let in (\ref{6.11}) the function }$c_{glob}\in G$\emph{\ and }%
\begin{equation*}
\left\Vert c_{glob}-c^{\ast }\right\Vert <\frac{\delta ^{3\mu }}{3}.
\end{equation*}%
\emph{Then there exists a} \emph{sufficiently small number }$\delta
_{0}=\delta _{0}\left( \Omega ,d,\omega ,z_{\zeta },a,\left\Vert
P\right\Vert _{H^{6}\left( Q_{T}\right) },\left\Vert \Phi \right\Vert
_{H^{5}\left( Q_{T}\right) },\mu \right) \in \left( 0,1\right) $\emph{\ such
that }$V_{\delta ^{3\mu }}\left( c^{\ast }\right) \subset G$,$\forall \delta
\in \left( 0,\delta _{0}\right) $\emph{\ and the functional }$E_{\alpha
}\left( c\right) $\emph{\ is strongly convex in }$V_{\delta ^{3\mu }}\left(
c^{\ast }\right) $\emph{\ with the strong convexity constant }$\alpha /4.$%
\emph{\ In other words, }
\begin{equation}
\left\Vert c_{1}-c_{2}\right\Vert ^{2}\leq \frac{2}{\delta ^{2\mu }}\left(
E_{\alpha }^{\prime }\left( c_{1}\right) -E_{\alpha }^{\prime }\left(
c_{2}\right) ,c_{1}-c_{2}\right) ,\text{ }\forall c_{1},c_{2}\in V_{\delta
^{3\mu }}\left( c^{\ast }\right) ,  \label{6.19}
\end{equation}%
\emph{where }$\left( ,\right) $\emph{\ is the scalar product in }$%
L_{2}\left( \Omega \right) $ \emph{and the Fr\'{e}chet derivative }$%
E_{\alpha }^{\prime }$\emph{\ is calculated via (\ref{6.15}).} \emph{%
Furthermore, there exists the unique regularized solution }$c_{\alpha \left(
\delta \right) }$\emph{, and }$c_{\alpha \left( \delta \right) }\in
V_{\delta ^{3\mu }/3}\left( x^{\ast }\right) .$\emph{\ In addition, the
gradient method of the minimization of the functional }$E_{\alpha }\left(
c\right) ,$\emph{\ which starts at }$c_{glob},$\emph{\ converges to }$%
c_{\alpha \left( \delta \right) }.$ \emph{Furthermore, let }$\xi \in \left(
0,1\right) $\emph{\ be an arbitrary number. Then there exists a number }$%
\delta _{1}=\delta _{1}\left( \Omega ,d,\omega ,z_{\zeta },a,\left\Vert
P\right\Vert _{H^{6}\left( Q_{T}\right) },\left\Vert \Phi \right\Vert
_{H^{5}\left( Q_{T}\right) },\mu ,\xi \right) \in \left( 0,\delta
_{0}\right) $\emph{\ such that for all }$\delta \in \left( 0,\delta
_{1}\right) $\emph{\ }
\begin{equation*}
\left\Vert c_{\alpha \left( \delta \right) }-c^{\ast }\right\Vert \leq
\left\{
\begin{array}{c}
\xi \left\Vert c_{glob}-c^{\ast }\right\Vert ,\text{ if }c_{glob}\neq
c^{\ast }, \\
\xi ,\text{ if }c_{glob}=c^{\ast }.%
\end{array}%
\right. \text{ }
\end{equation*}%
\emph{In other words, the regularized solution }$c_{\alpha \left( \delta
\right) }$\emph{\ is more accurate than the solution obtained on the first
stage of our two-stage numerical procedure}. \emph{Furthermore, since }$%
E_{\alpha \left( \delta \right) }^{\prime }\left( c_{\alpha \left( \delta
\right) }\right) =0,$\emph{\ then (\ref{6.19}) implies that}%
\begin{equation*}
\left\Vert c-c_{\alpha \left( \delta \right) }\right\Vert \leq \frac{2}{%
\delta ^{2\mu }}\left\Vert E_{\alpha \left( \delta \right) }^{\prime }\left(
c\right) \right\Vert _{L_{2}\left( \Omega \right) },\forall c\in V_{\delta
^{3\mu }}\left( c^{\ast }\right) .
\end{equation*}%
\emph{\ }

Theorem 6.4 follows from Theorems 5.1 and 6.3 as well as from Theorem 4.11.3
of \cite{BK1}.

\textbf{Theorem 6.4.} \emph{Let conditions of Theorem 6.3 hold. Let }$%
\left\Vert c^{\ast }\right\Vert \leq B,$\emph{\ where the constant }$B$\emph{%
\ is given. Let }$M_{n}\subset H$\emph{\ be the subspace obtained after }$n$%
\emph{\ mesh refinements as described in section 4. Let }$h_{n}$\emph{\ be
the maximal grid step size of the subspace }$M_{n}$.\emph{\ Let }$D=D\left(
\Omega ,a,d,\omega ,z_{\zeta }\right) >0$\emph{\ be the constant of Theorem
6.1 and }$K$\emph{\ be the constant in (\ref{4.6}). There exists a constant }%
$\overline{N}_{2}=\overline{N}_{2}\left( D,T,\left\Vert P\right\Vert
_{H^{6}\left( Q_{T}\right) },\left\Vert \Phi \right\Vert _{H^{5}\left(
Q_{T}\right) }\right) $\emph{\ such that if }
\begin{equation*}
h_{n}\leq \frac{\delta ^{4\mu }}{5B\overline{N}_{2}K},
\end{equation*}%
\emph{then there exists the unique minimizer }$c_{n}$ \emph{of the
functional (\ref{6.11}) on the set }$G\cap M_{n}$\emph{. Furthermore, }$%
c_{n}\in V_{\delta ^{3\mu }}\left( x^{\ast }\right) \cap M_{n}$\emph{\ and
the following a posteriori error estimate holds}
\begin{equation}
\left\Vert c_{n}-c_{\alpha \left( \delta \right) }\right\Vert \leq \frac{2}{%
\delta ^{2\mu }}\left\Vert E_{\alpha \left( \delta \right) }^{\prime }\left(
c_{n}\right) \right\Vert _{L_{2}\left( \Omega \right) }.  \label{6.21}
\end{equation}

The estimate (\ref{6.21}) is \emph{a posteriori} because it is obtained
after the function $c_{n}$ is calculated. Theorem 6.5 follows from Theorems
5.2, 5.3, 6.4, also see Theorem 4.11.4 in \cite{BK1}.

\textbf{Theorem 6.5 }(relaxation)\textbf{. }\emph{Assume that conditions of
Theorem 6.4 hold. Let }$c_{n}\in V_{\delta ^{3\mu }}\left( x^{\ast }\right)
\cap M_{n}$\emph{\ be the unique minimizer of the Tikhonov functional (\ref%
{6.11}) on the set }$G\cap M_{n}$ \emph{(Theorem 6.4).} \emph{Assume that
the regularized solution }$c_{\alpha \left( \delta \right) }\neq c_{n},$%
\emph{\ i.e. }$c_{\alpha \left( \delta \right) }\notin M_{n}.$\emph{\ Let }$%
\eta \in \left( 0,1\right) $\emph{\ be an arbitrary number. Then one can
choose the maximal grid size }$h_{n+1}=h_{n+1}\left( B,\overline{N}%
_{2},K,\delta ,z_{\zeta },\mu ,\eta \right) \in \left( 0,h_{n}\right] $\emph{%
\ of the mesh refinement number }$\left( n+1\right) $\emph{\ so small that }
\begin{equation}
\left\Vert c_{n+1}-c_{\alpha \left( \delta \right) }\right\Vert \leq \eta
\left\Vert c_{n}-c_{\alpha \left( \delta \right) }\right\Vert ,  \label{6.22}
\end{equation}%
\emph{where the number }$\overline{N}_{2}$\emph{\ was defined in Theorem
6.4. Let }$\xi \in \left( 0,1\right) $\emph{\ be an arbitrary number. Then
there exists a sufficiently small number }$\delta _{0}=\delta _{0}\left( A,%
\overline{N}_{2},K,\delta ,z_{\zeta },\xi ,\mu ,\eta \right) \in \left(
0,1\right) $\emph{\ and a decreasing sequence of maximal grid step sizes }$%
\left\{ h_{k}\right\} _{k=1}^{n+1},h_{k}=h_{k}\left( B,\overline{N}%
_{2},K,\delta ,z_{\zeta },\xi ,\mu .\eta \right) $ \emph{such that if }$%
\delta \in \left( 0,\delta _{0}\right) ,$ \emph{then}
\begin{equation}
\left\Vert c_{k+1}-c^{\ast }\right\Vert \leq \eta ^{k}\left\Vert
c_{1}-c_{\alpha \left( \delta \right) }\right\Vert +\left\{
\begin{array}{c}
\xi \left\Vert c_{glob}-c^{\ast }\right\Vert ,\text{ if }c_{glob}\neq
c^{\ast }, \\
\xi ,\text{ if }c_{glob}=c^{\ast },%
\end{array}%
\right. ,k=1,...,n.  \label{6.23}
\end{equation}

Theorem 6.6 follows from Theorems 5.4 and 6.5.

\textbf{Theorem 6.6}. (relaxation for local mesh refinements). \emph{Assume
that conditions of Theorem 6.5 hold. Let }$\Omega =\Omega _{1}\cup \Omega
_{2}.$\emph{\ Suppose that mesh refinements are performed only in the
subdomain }$\Omega _{2}.$\emph{\ Let }$h^{\left( 1\right) }$\emph{\ be the
maximal grid step size in }$\Omega _{1}.$\emph{\ Then there exists a
sufficiently small number }$\delta _{0}=\delta _{0}\left( B,\overline{N}%
_{2},K,z_{\zeta },\mu ,\eta \right) \in \left( 0,1\right) $\emph{\ and a
decreasing sequence of maximal grid step sizes }$\left\{ \widetilde{h}%
_{k}\right\} _{k=1}^{n+1},\widetilde{h}_{k}=\widetilde{h}_{k}$\emph{\ }$%
\left( B,\overline{N}_{3},K,z_{\zeta },\mu ,\eta \right) $\emph{\ of meshes
in }$\Omega _{2}$\emph{\ such that if }$\left\Vert \nabla c_{\alpha \left(
\delta \right) }\right\Vert _{L_{\infty }\left( \Omega _{1}\right) }$\emph{\
is so small that if }%
\begin{equation*}
\frac{2K\overline{N}_{3}}{\delta ^{2\mu }}\left\Vert \nabla c_{\alpha \left(
\delta \right) }\right\Vert _{L_{\infty }\left( \Omega _{1}\right)
}h^{\left( 1\right) }\leq \frac{\eta }{2}\left\Vert c_{k}-c_{\alpha \left(
\delta \right) }\right\Vert ,k=1,...,n\text{ and }\delta \in \left( 0,\delta
_{0}\right) ,
\end{equation*}
\emph{then (\ref{6.23}) holds with the replacement of }$\left\{
h_{k}\right\} _{k=1}^{n+1}$\emph{\ with }$\left\{ \widetilde{h}_{k}\right\}
_{k=1}^{n+1}.$

\emph{Here the number} $\overline{N}_{3}=\overline{N}_{3}\left(
D,T,\left\Vert P\right\Vert _{H^{6}\left( Q_{T}\right) },\left\Vert \Phi
\right\Vert _{H^{5}\left( Q_{T}\right) }\right) >0$.

\section{Mesh Refinement Recommendations and the Adaptive Algorithm}

\label{sec:7}

\subsection{Mesh Refinement Recommendations}

\label{sec:7.1}

Recommendations for mesh refinements are based on the theory of section 6.
We now present some partly rigorous and partly heuristic considerations
which lead to these recommendations. The latter means that both mesh
refinement recommendations listed below should be verified numerically. We
come back to the arguments presented in the paragraph above Theorem 5.4. To
simplify the presentation, assume, for example that
\begin{equation}
\nabla c_{\alpha \left( \delta \right) }\left( x\right) =\nabla c^{\ast
}\left( x\right) =0\text{ for }x\in \Omega _{1}.  \label{7.0}
\end{equation}%
A more general case when functions $c_{\alpha \left( \delta \right) }\left(
x\right) ,c^{\ast }\left( x\right) $ change slowly in $\Omega _{1}$ can be
considered similarly. Using (\ref{4.5}) and (\ref{7.0}), we obtain that $%
\left( c_{\alpha \left( \delta \right) }-P_{k}c_{\alpha \left( \delta
\right) }\right) \left( x\right) =0$ for $x\in \Omega _{1},\forall k\geq 1.$
Hence, by (\ref{4.5})
\begin{equation*}
\left\Vert c_{\alpha \left( \delta \right) }-P_{n+1}c_{\alpha \left( \delta
\right) }\right\Vert _{L_{2}\left( \Omega \right) }=\left\Vert c_{\alpha
\left( \delta \right) }-P_{n+1}c_{\alpha \left( \delta \right) }\right\Vert
_{L_{2}\left( \Omega _{2}\right) }\leq K\left\Vert \nabla c_{\alpha \left(
\delta \right) }\right\Vert _{L_{\infty }\left( \Omega _{2}\right) }%
\widetilde{h}_{n+1},
\end{equation*}%
where $\widetilde{h}_{n+1}$ is the maximal grid step size in $\Omega _{2}$
after $n+1$ mesh refinements. Hence, using the second equality (\ref{5.12})
and (\ref{5.14}), we obtain%
\begin{equation}\label{7.2_1}
\left\Vert c_{n+1}-c_{\alpha \left( \delta \right) }\right\Vert \leq \frac{2K%
}{\delta ^{2\mu }}\left\Vert E_{\alpha \left( \delta \right) }^{\prime
}\left( c_{n+1}\right) \right\Vert \left\Vert \nabla c_{\alpha \left( \delta
\right) }\right\Vert _{L_{\infty }\left( \Omega _{2}\right) }\widetilde{h}%
_{n+1}.
\end{equation}%
Given a function $f\in C\left( \overline{\Omega }\right) ,$ the main impact
in the norm $\left\Vert f\right\Vert _{L_{2}\left( \Omega \right) }$ is
provided by neighborhoods of those points $x\in \overline{\Omega }$ \ where
the function $\left\vert f\left( x\right) \right\vert $ achieves its maximal
value. Hence, (\ref{7.2_1}) indicates that we should decrease the maximal grid
step size $\widetilde{h}_{n+1}$ (i.e. refine mesh) in neighborhoods of those
points $x\in \Omega _{2}$ where the function $\left\vert E_{\alpha }^{\prime
}\left( c_{n+1}\right) \left( x\right) \right\vert $ achieves its maximal
values, where the function $E_{\alpha \left( \delta \right) }^{\prime
}\left( c_{n+1}\right) \left( x\right) \in C\left( \overline{\Omega }\right)
$ is given by formula (\ref{6.15}). Although after $n$ mesh refinements we
know only the function $c_{n}\in M_{n}$ rather than the function $c_{n+1}\in
M_{n+1}$, still, since functions $c_{n}$ and $c_{n+1}$ are sufficiently
close to each other, we should likely refine mesh in neighborhoods of those
points $x\in \Omega _{2}$ where the function $\left\vert E_{\alpha }^{\prime
}\left( c_{n}\right) \left( x\right) \right\vert $ achieves its maximal
values. These considerations lead to two mesh refinement recommendations
below.

\textbf{The} \textbf{First Mesh Refinement Recommendation.} \emph{Let }$%
\beta _{1}\in \left( 0,1\right) $\emph{\ be the tolerance number. Refine the
mesh in such subdomains of \ }$\Omega _{2}$\emph{\ where}
\begin{equation}
\left\vert E_{\alpha }^{\prime }\left( c_{n}\right) \left( x\right)
\right\vert \geq \beta _{1}\max_{\overline{\Omega }_{2}}\left\vert E_{\alpha
}^{\prime }\left( c_{n}\right) \left( x\right) \right\vert .  \label{7.2}
\end{equation}

To figure out the second mesh refinement recommendation, we note that by (%
\ref{6.15}) and (\ref{6.16})
\begin{equation*}
\left\vert E_{\alpha \left( \delta \right) }^{\prime }\left( c_{n}\right)
\left( x\right) \right\vert \leq \alpha \left( \left\Vert c_{n}\right\Vert
_{C\left( \overline{\Omega }\right) }+\left\Vert c_{glob}\right\Vert
_{C\left( \overline{\Omega }\right) }\right) +\left\Vert c_{n}\right\Vert
_{C\left( \overline{\Omega }\right) }^{2}\exp \left( DT\right) \left(
\left\Vert P\right\Vert _{H^{6}\left( Q_{T}\right) }^{2}+\left\Vert \Phi
\right\Vert _{H^{5}\left( Q_{T}\right) }^{2}\right) .
\end{equation*}%
Since $\alpha $ is small, then the second term in the right hand side of
this estimate dominates. Next, since we have decided to refine the mesh in
neighborhoods of those points, which deliver maximal values for the function
$\left\vert E_{\alpha \left( \delta \right) }^{\prime }\left( c_{n}\right)
\left( x\right) \right\vert ,$ then we obtain the following mesh refinement
recommendation.

\textbf{Second Mesh Refinement Recommendation.} \emph{Let }$\beta _{2}\in
\left( 0,1\right) $\ \emph{be the tolerance number. Refine the mesh in such
subdomains of }$\Omega _{2}$\emph{\ where}
\begin{equation}
c_{n}\left( x\right) \geq \beta _{2}\max_{\overline{\Omega }_{2}}c_{n}\left(
x\right) ,  \label{7.3}
\end{equation}

In fact, these two mesh refinement recommendations do not guarantee of
course that the minimizer obtained on the corresponding finer mesh would be
indeed more accurate than the one obtained on the coarser mesh. This is
because right hand sides of formulas (\ref{7.2}) and (\ref{7.3}) are
indicators only. Thus, numerical verifications are necessary. As to
tolerance numbers $\beta _{1}$ and $\beta _{2},$ they should be chosen
numerically. Indeed, if we would choose $\beta _{1},\beta _{2}\approx 1,$
then we would refine the mesh in too narrow regions. On the other hand, if
we would choose $\beta _{1},\beta _{2}\approx 0,$ then we would refine the
mesh in almost the entire subdomain $\Omega _{2},$ which is inefficient.\

\subsection{The adaptive algorithm}

\label{sec:7.2}

Since this algorithm was described in detail in a number of publications,
see, e.g. \cite{BK1,BK3}, we outline it only briefly here. Recall that the
adaptivity is used on the second stage of our two-stage numerical procedure
(section 1). On the first stage the approximately globally convergent
algorithm is applied. It was proven, within the framework of the so-called
Second Approximate Mathematical Model, that this algorithm delivers a good
approximation for the exact solution $c^{\ast }\left( x\right) $ of the
above CIP, see Theorem 2.9.4 in \cite{BK1} as well as Theorem 5.1 in \cite%
{BK5}. We start the adaptivity on the same mesh on which the algorithm of
the first stage has worked. In our experience, this mesh does not provide an
improvement of the image. On each mesh we find an approximate solution of
the equation $E_{\alpha }^{\prime }\left( c\right) =0.$ Hence, by (\ref{6.15}%
)\emph{\ }we find an approximate solution of the following equation on each
mesh
\begin{equation*}
\alpha \left( c-c_{glob}\right) \left( x\right) -\int\limits_{0}^{T}\left(
u_{t}\lambda _{t}\right) \left( x,t\right) ~dt=0.
\end{equation*}%
For each newly refined mesh we first linearly interpolate the function $%
c_{glob}\left( x\right) $ on\ it. Since this function was initially computed
as a linear combination of finite elements forming the initial mesh and
since all our finite elements are piecewise linear functions, then
subsequent linear interpolations on finer meshes do not change the function $%
c_{glob}\left( x\right) $. On each mesh we iteratively update approximations
$c_{\alpha }^{n}$ of the function $c_{\alpha \left( \delta \right) }$. To do
this, we use the quasi-Newton method with the classic BFGS update formula
with the limited storage \cite{Nocedal}. Denote
\begin{equation*}
\varphi ^{n}(x)=\alpha (c_{\alpha }^{n}-c_{glob})\left( x\right)
-\int_{0}^{T}\left( v_{ht}\lambda _{ht}\right) \left( x,t,c_{\alpha
}^{n}\right) dt,
\end{equation*}%
where functions $v_{h}\left( x,t,c_{\alpha }^{n}\right) ,\lambda _{h}\left(
x,t,c_{\alpha }^{n}\right) $\ are FEM solutions of state and adjoint
problems (\ref{6.9}), (\ref{6.10}) with $c:=c_{\alpha }^{n}$. We stop
computing $c_{\alpha }^{n}$ if either $||\varphi ^{n}||_{L_{2}(\Omega )}\leq
10^{-5}$ or norms $||\varphi ^{n}||_{L_{2}(\Omega )}$ are stabilized. Of
course, only discrete norms are considered here.

For a given mesh obtained after $n$ mesh refinements, let $c_{n}$ be
the last computed function on which we have stopped. Next, we compute
the function $\left\vert E_{\alpha }^{\prime }\left( c_{n}\right)
\left( x\right) \right\vert $ using (\ref{6.15}), where
$v:=v_{h}\left( x,t,c_{n}\right) ,\lambda :=\lambda _{h}\left(
x,t,c_{n}\right) .$ If we use both above mesh refinement
recommendations, then we refine the mesh in neighborhoods of all grid
points satisfying (\ref{7.2}) and (\ref{7.3}). In some studies,
however, we use only the first recommendation. In this case we refine
the mesh in neighborhoods of all grid points satisfying only
(\ref{7.2}).

\section{\textbf{Numerical Studies}}

\label{sec:8}

We present here three numerical examples of the performance of our two-stage
numerical procedure: one for computationally simulated and two for
experimental data. More numerical tests of the adaptivity technique can be
found in \cite{AB,BJ1,Beil,BJ2,BC,Beilina1,BK1,BK2,BK3,BK4,BKK}. In Test 1
we have used only the First Mesh Refinement Recommendation, and in Tests 2,3
we have used both recommendations. Since the numerical method of the first
stage of our procedure is not a focus of this paper, and since it was
described earlier in, e.g. \cite{BK1,BK2,BK3,BK4,KFBPS,KBK,KBKSNF}, we do
not describe it here.

\subsection{Computationally simulated data}

\label{sec:8.1}

\begin{figure}[tbp]
\begin{center}
\begin{tabular}{ccc}
{\includegraphics[scale=0.3,clip=]{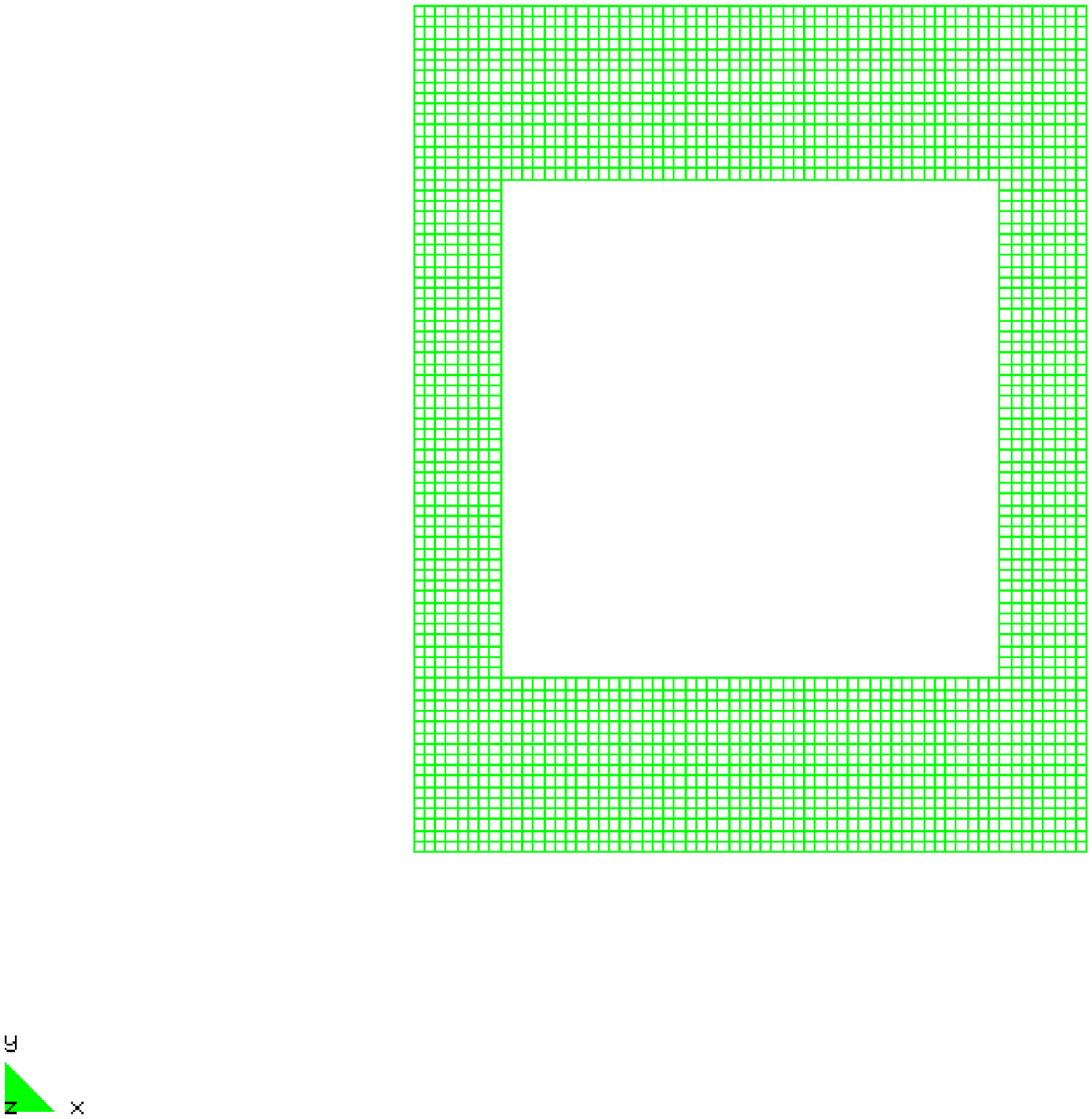}} & {%
\includegraphics[scale=0.2,clip=]{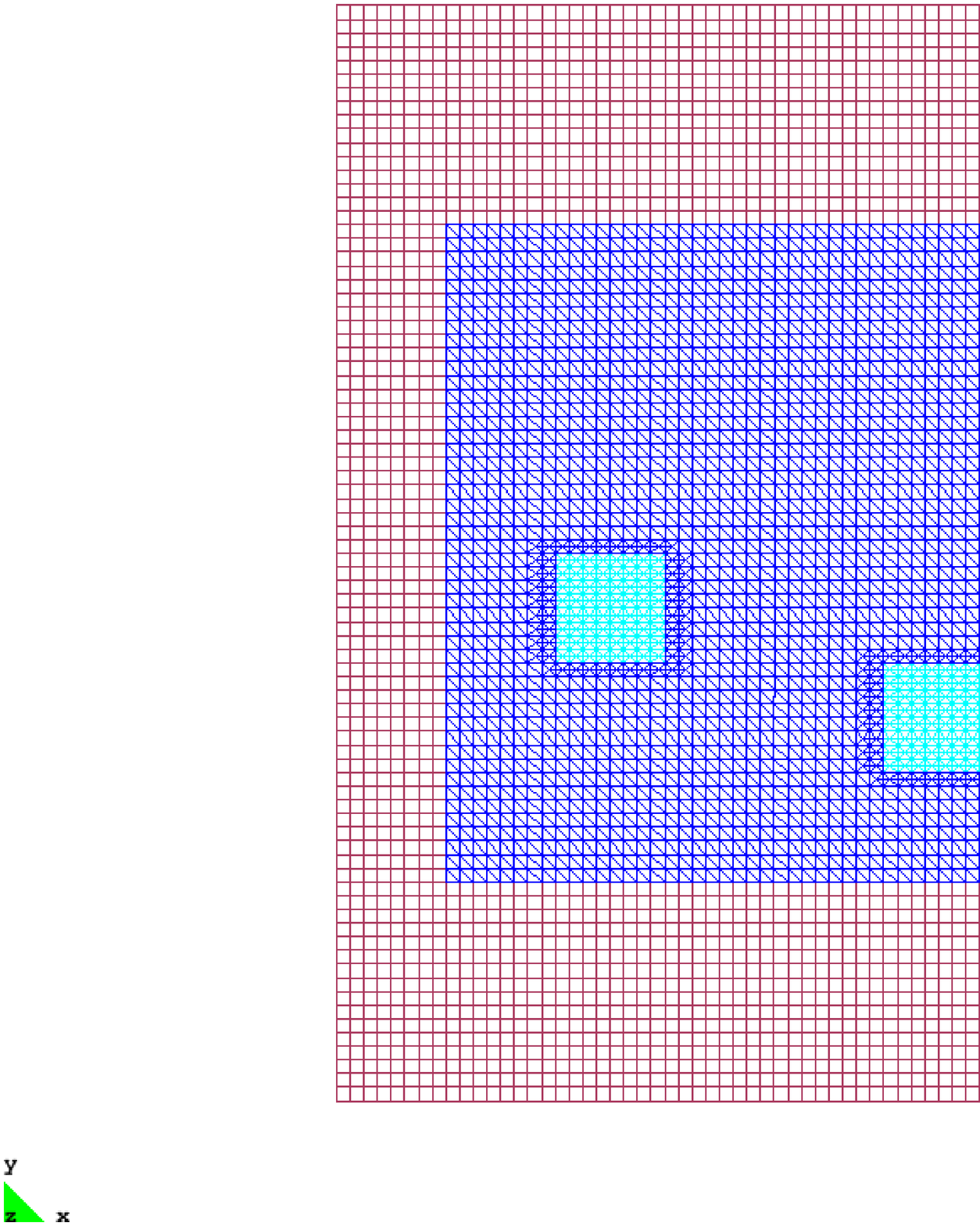}} & {%
\includegraphics[scale=0.15,clip=]{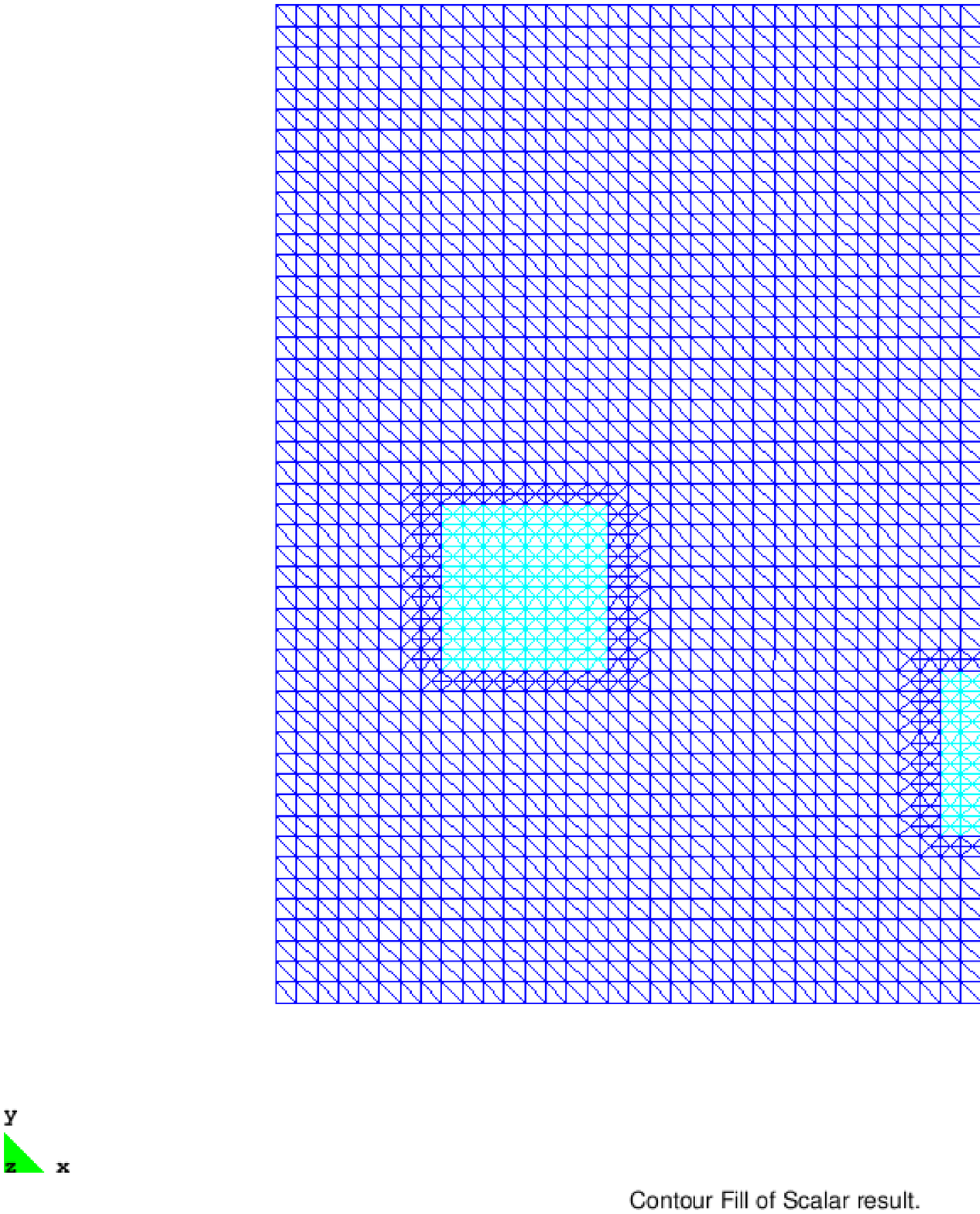}} \\
(a) $G_{FDM}$ & (b) $G = G_{FEM} \cup G_{FDM}$ & (c) $G_{FEM}=\Omega$%
\end{tabular}
\end{center}
\caption{The hybrid mesh (b) is a combinations of a structured mesh (a),
where FDM is applied, and a mesh (c), where we use FEM, with a thin
overlapping of structured elements. The solution of the inverse problem is
computed in the square $\Omega$ and $c(x)=1$ for $x \in G\diagdown\Omega$.}
\label{fig:Figure1}
\end{figure}

\textbf{Test 1}.  We conducted
computational simulations in two dimensions. Since it is impossible to
computationally solve equation (\ref{6.2}) in the entire space
$\mathbb{R}^{2},$ we have conducted data simulations in the rectangle
$G=\left[ -4,4\right] \times \left[ -5,5\right] .$ To simulate the
boundary data $g\left( x,t\right) $, we have solved the forward
problem by the hybrid FEM/FDM method \cite{BSA} using the software
package WavES \cite{W}.  To do this, we split the domain $G$ in
two subdomains $G=G_{FEM}\cup G_{FDM}, $ see Figure \ref{fig:Figure1}. Here
$G_{FEM}:=\Omega =\left[ -3,3\right] \times \left[ -3,3\right] $ and
$G_{FDM}=G\diagdown G_{FEM}.$
 The coefficient $c(x)$ is unknown in the domain
    $\Omega \subset G$ and is defined as
\begin{equation}
c(x) =\left\{
\begin{array}{ll}
1 & \text{ in }G_{FDM}, \\
1+b(x) &\text{ in } G_{FEM}, \\
4 &\text{ in small squares of Figure \ref{fig:Figure1}},
\end{array}
\right.  \label{Y}
\end{equation}
where the function $b(x) \in \Omega_{FEM}$ is defined as
\begin{equation}
b(x) =\left\{
\begin{array}{ll}
0 &\text{for} (x_1,x_2) \in  \Omega_{FEM}: -2.875 < x_1 < 0, -2.875 <  x_2 < 0, \\
0.5\sin ^{2}\left( \frac{\pi x_{1}}{2.875}\right) \sin ^{2}\left( \frac{\pi
x_{2}}{2.875}\right) &  \text{ otherwise.}\\
\end{array}
\right.  \notag
\end{equation}

The spatial mesh consists of triangles in $G_{FEM}$ and of squares in
$G_{FDM}$ with the grid step size $\overline{h}=0.125$ both in
overlapping regions and in $G_{FDM}.$ There is no reason to refine
mesh in $G_{FDM}$ since $c\left( x\right) =1$ in $G_{FDM}.$ Let
$\partial G_{1}$ and $\partial G_{2}$ be, respectively, top and bottom
sides of the rectangle $G$ and $\partial G_{3}$ be the union of
vertical sides of $G$. We use first order absorbing boundary
conditions on $\partial G_1 \cup \partial G_{2}$ \cite{EM} and zero
Neumann boundary condition on $\partial G_{3}.$

Let $\overline{s}$ be the upper value of the Laplace transform of the
solution of our forward problem. We use this transform on the first stage of
our two-stage numerical procedure. It was found that for the above domain $%
\Omega $ the optimal value is $\overline{s}=7.45.$ Consider the function $%
f\left( t\right) ,$%
\begin{equation*}
f\left( t\right) =\left\{
\begin{array}{c}
0.1\left[ \sin \left( \overline{s}t-\pi /2\right) +1\right] ,t\in \left[
0,t_{1}\right] ,t_{1}=2\pi /\overline{s}, \\
0,t\in \left( t_{1},T\right] ,T=17.8t_{1}.%
\end{array}%
\right.
\end{equation*}%
The forward problem for data simulations is
\begin{equation}
\begin{split}
c\left( x\right) u_{tt}-\Delta u& =0,~~~\mbox{in}~G\times (0,T), \\
u(x,0)& =0,~u_{t}(x,0)=0,~\mbox{in}~G, \\
\partial _{n}u\big \vert_{\partial G_{1}}& =f\left( t\right) ,~\mbox{on}%
~\partial G_{1}\times (0,t_{1}], \\
\partial _{n}u\big \vert_{\partial G_{1}}& =-\partial _{t}u,~\mbox{on}%
~\partial G_{1}\times (t_{1},T), \\
\partial _{n}u\big \vert_{\partial G_{2}}& =-\partial _{t}u,~\mbox{on}%
~\partial G_{2}\times (0,T), \\
\partial _{n}u\big \vert_{\partial G_{3}}& =0,~\mbox{on}~\partial
G_{3}\times (0,T).
\end{split}
\label{8.1}
\end{equation}%
The solution of this problem gives us the function $g\left( x,t\right)
=u\mid _{S_{T}}.$ Next, the coefficient $c\left( x\right) $ is
\textquotedblleft forgotten" and we apply the two-stage numerical procedure
to reconstruct it from the function $g\left( x,t\right) .$ To have noisy
data, we have added the random noise to the function $g\left( x,t\right) $
as
\begin{equation}\label{noise}
g_{i,j}=g\left( x^{i},t^{j}\right) \left[ 1+0.02\alpha _{j}\left( g_{\max
}-g_{\min }\right) \right] .
\end{equation}
Here $x^{i}\in \partial \Omega $ and $t^{j}\in \left[ 0,T\right] $ are mesh
points on $\partial \Omega $ and $\left[ 0,T\right] $ respectively, $g_{\min
}$ and $g_{\max }$ are minimal and maximal values of the function $g$ and $%
\alpha _{j}\in \left[ -1,1\right] $ is the random variable.
The ``inverse crime" was not committed
here since we have introduced the noise in the data and because the grids in
both stages of our two-stage numerical procedure were different from the one
which was used to solve the problem (\ref{8.1}).

\begin{figure}[tbp]
\begin{center}
\begin{tabular}{cc}
{\includegraphics[scale=0.25,clip=]{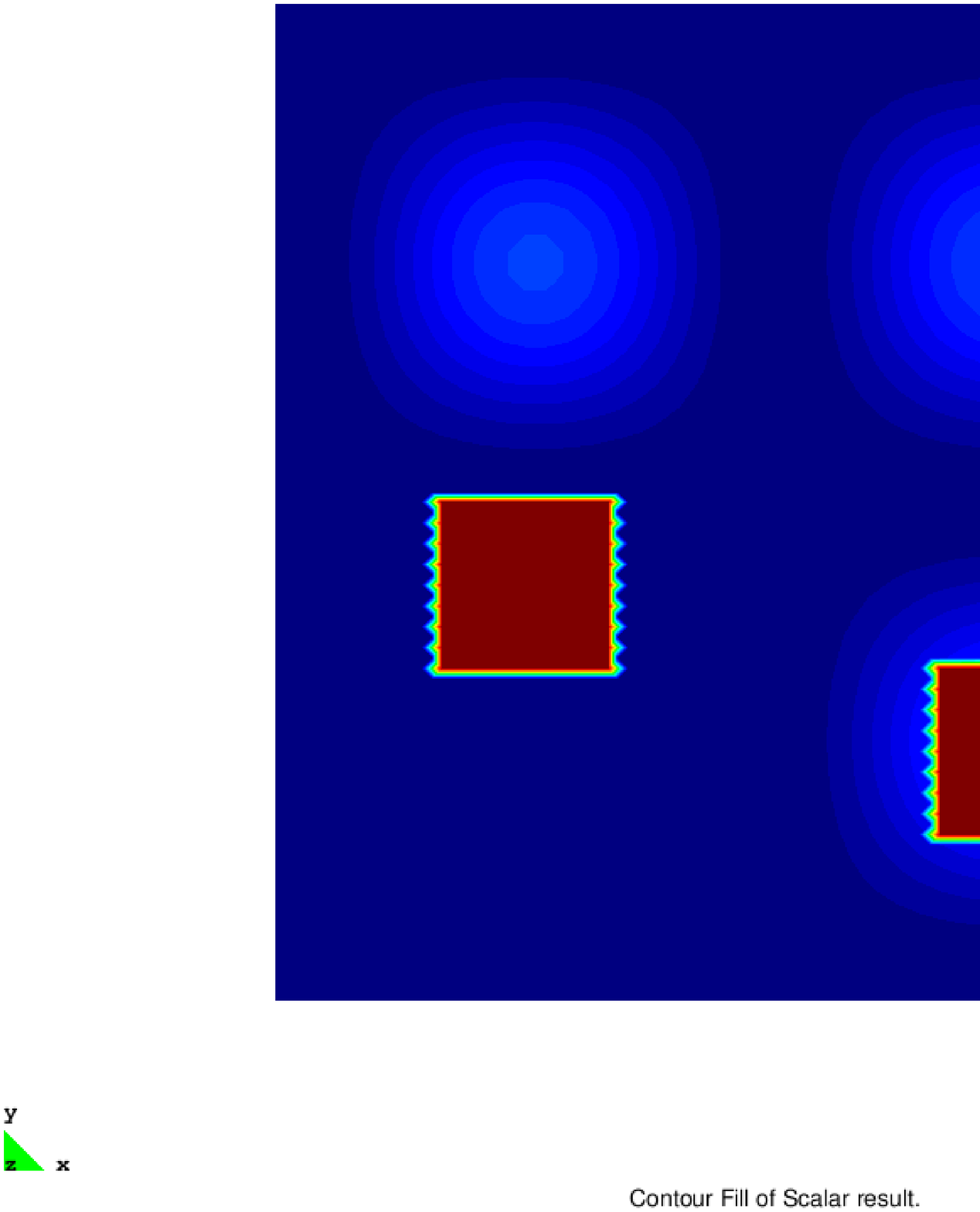}} &
{\includegraphics[scale=0.27,clip=]{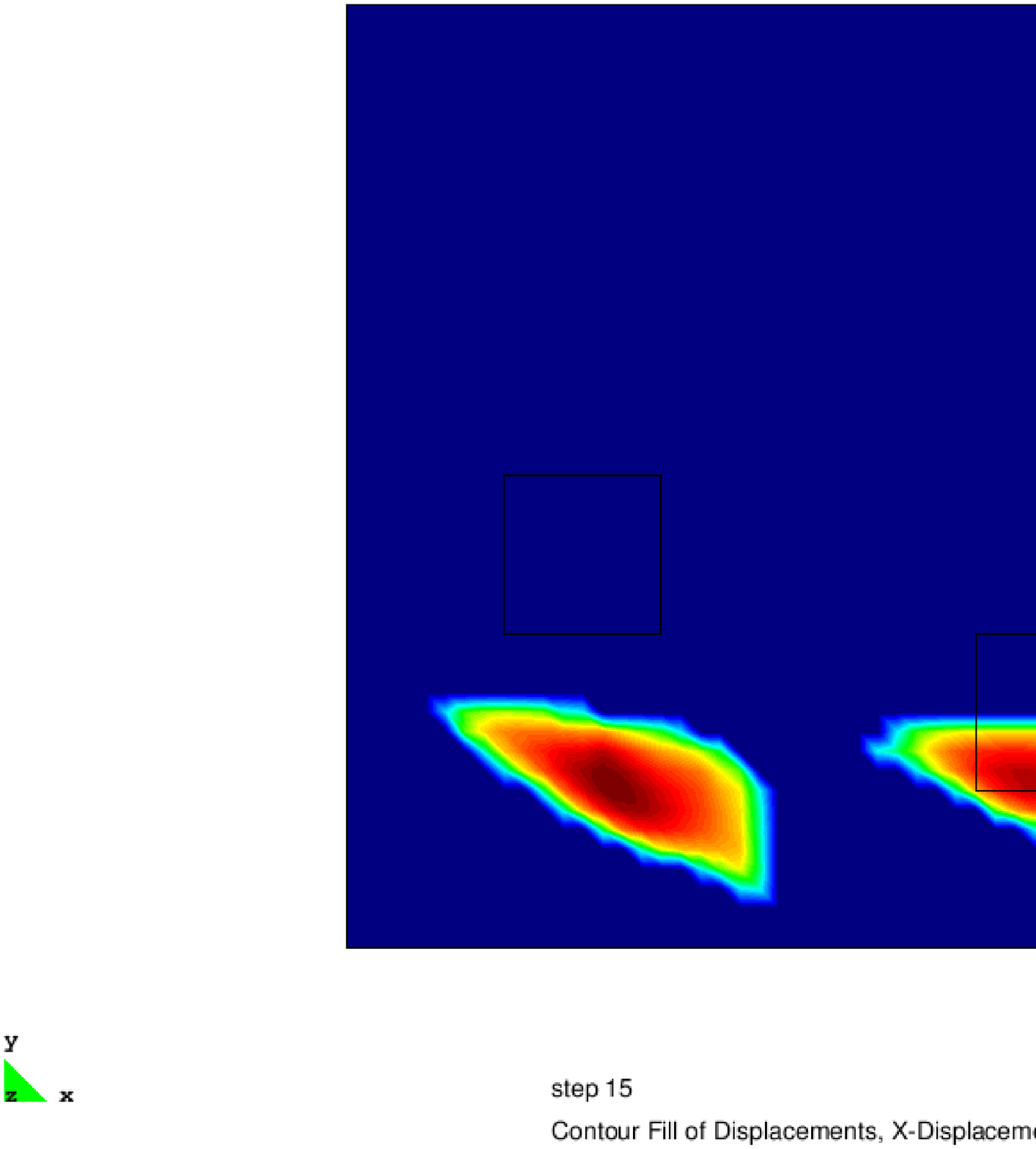}} \\
a)  exact coefficient  $c(x)$ & b)  Coefficient $c_{glob}$ reconstructed on the
first stage\\
\end{tabular}
\caption{ a) Spatial distribution of the exact coefficient $c(x)$.
b) Result of the performance of the
approximately globally convergent algorithm (first stage). The spatial
distribution of the computed coefficient $c_{glob}$ displayed. Here $\max
c_{glob}\left( x\right) =3.2,$ whereas $\max c\left( x\right) =4.$ Hence, we
have 20\% error in imaging of the maximal value of the function $c\left(
x\right) .$ The slowly changing part of the function $c\left( x\right) $,
i.e. the second raw in the above definition of the function $b\left(
x\right) ,$ is not imaged. Comparison with Figure a) shows that while the
location of the right inclusion is imaged correctly, the left one still
needs to be moved upwards. This is done on the second stage of our two-stage
numerical procedure, i.e. on the adaptivity stage. On this stage we take the
function $c_{glob}\left( x\right) $ as the starting point for the
minimization of the Tikhonov functional (\ref{6.11}).\ The second stage refines
the image of the first.
}
\label{fig:Figure2}
\end{center}
\end{figure}

\emph{1. The approximately globally convergent stage}. Since we focus on the
adaptivity in this paper, we do not describe this algorithm here and refer
to section 2.6.1 of \cite{BK1} instead. Figure \ref{fig:Figure2} displays the result of this
stage.

\begin{figure}[tbp]
\begin{center}
\begin{tabular}{ccc}
{\includegraphics[scale=0.18,clip=]{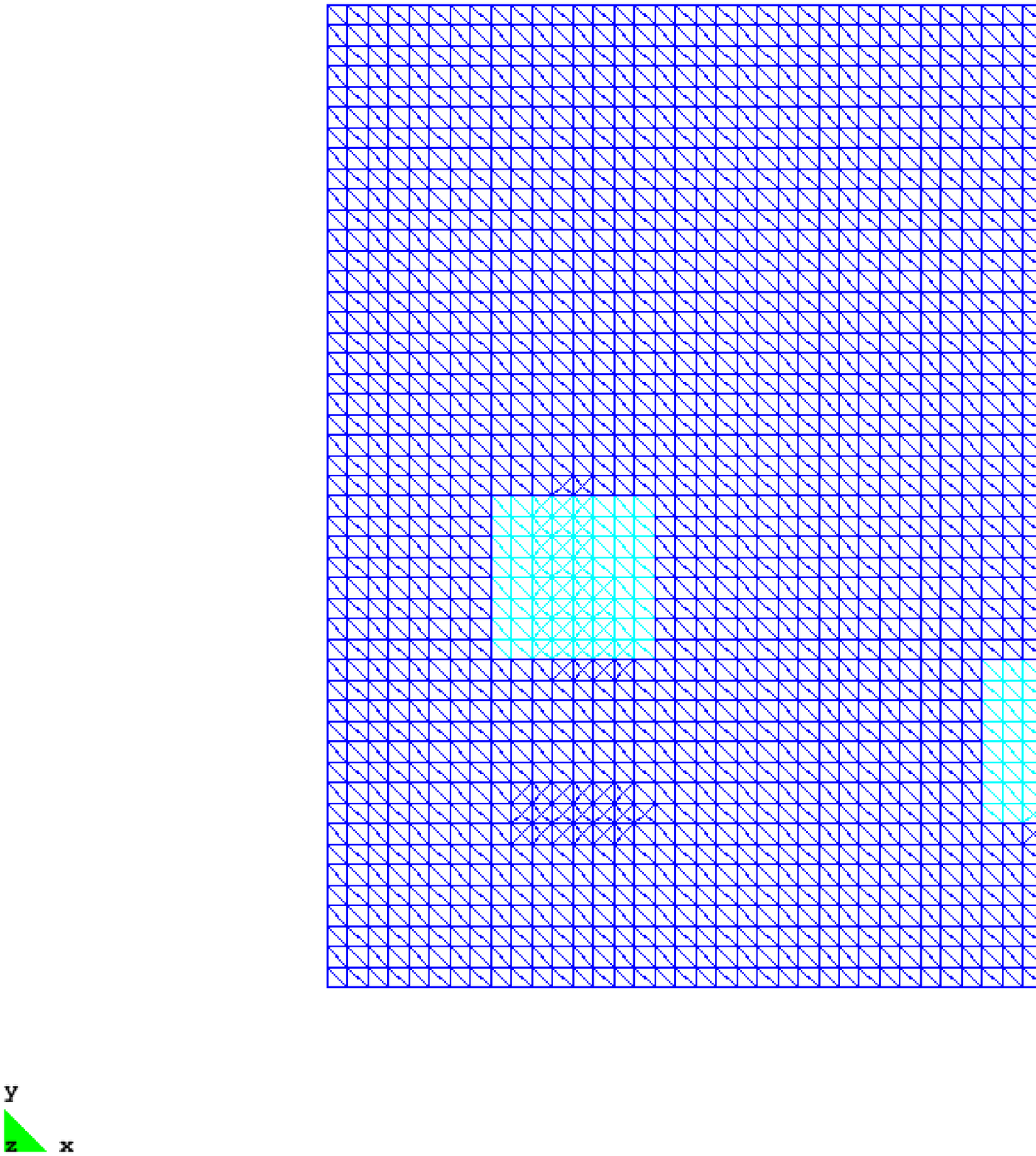}} &{
\includegraphics[scale=0.2,clip=]{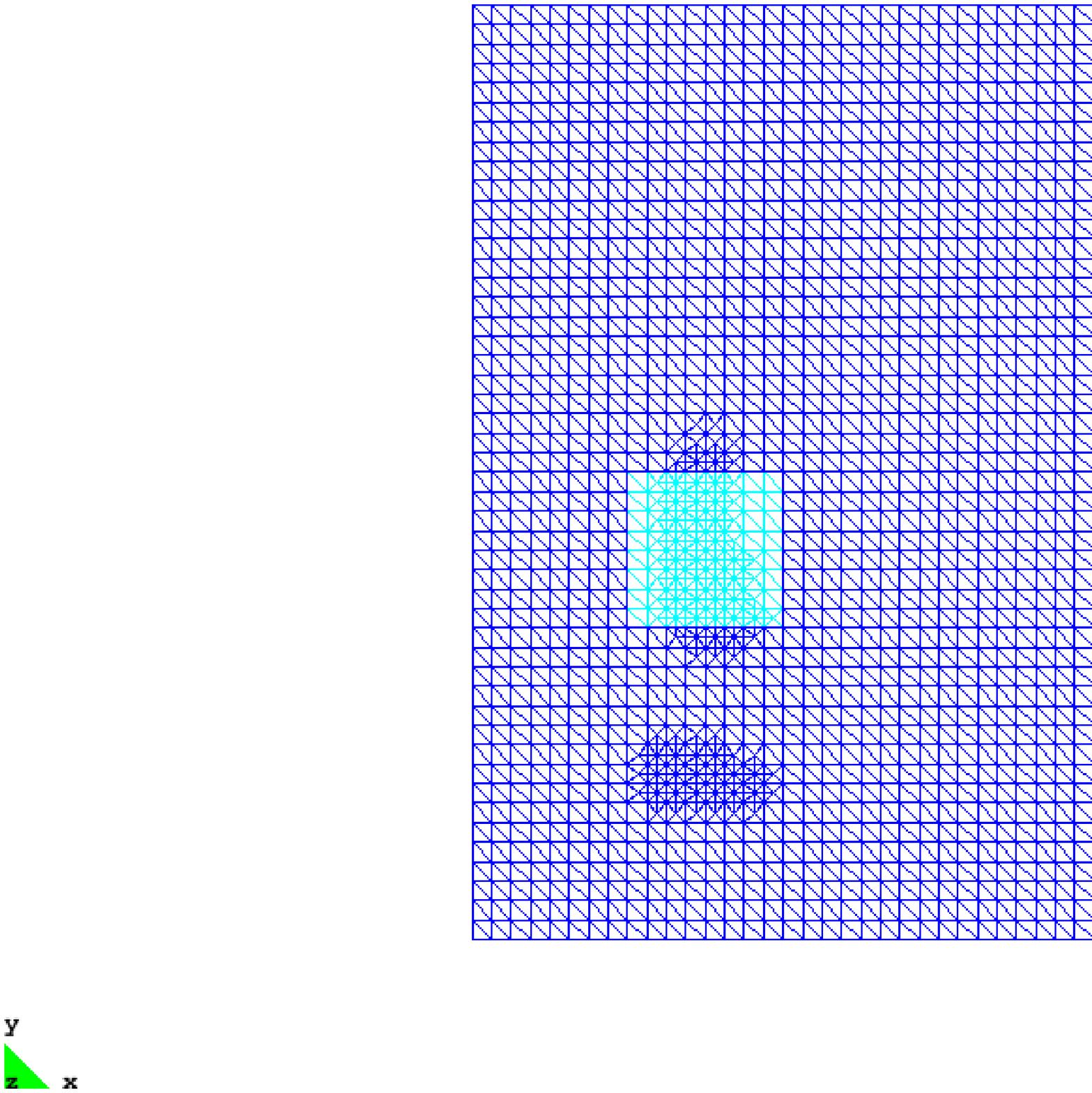}} & {
\includegraphics[scale=0.2,clip=]{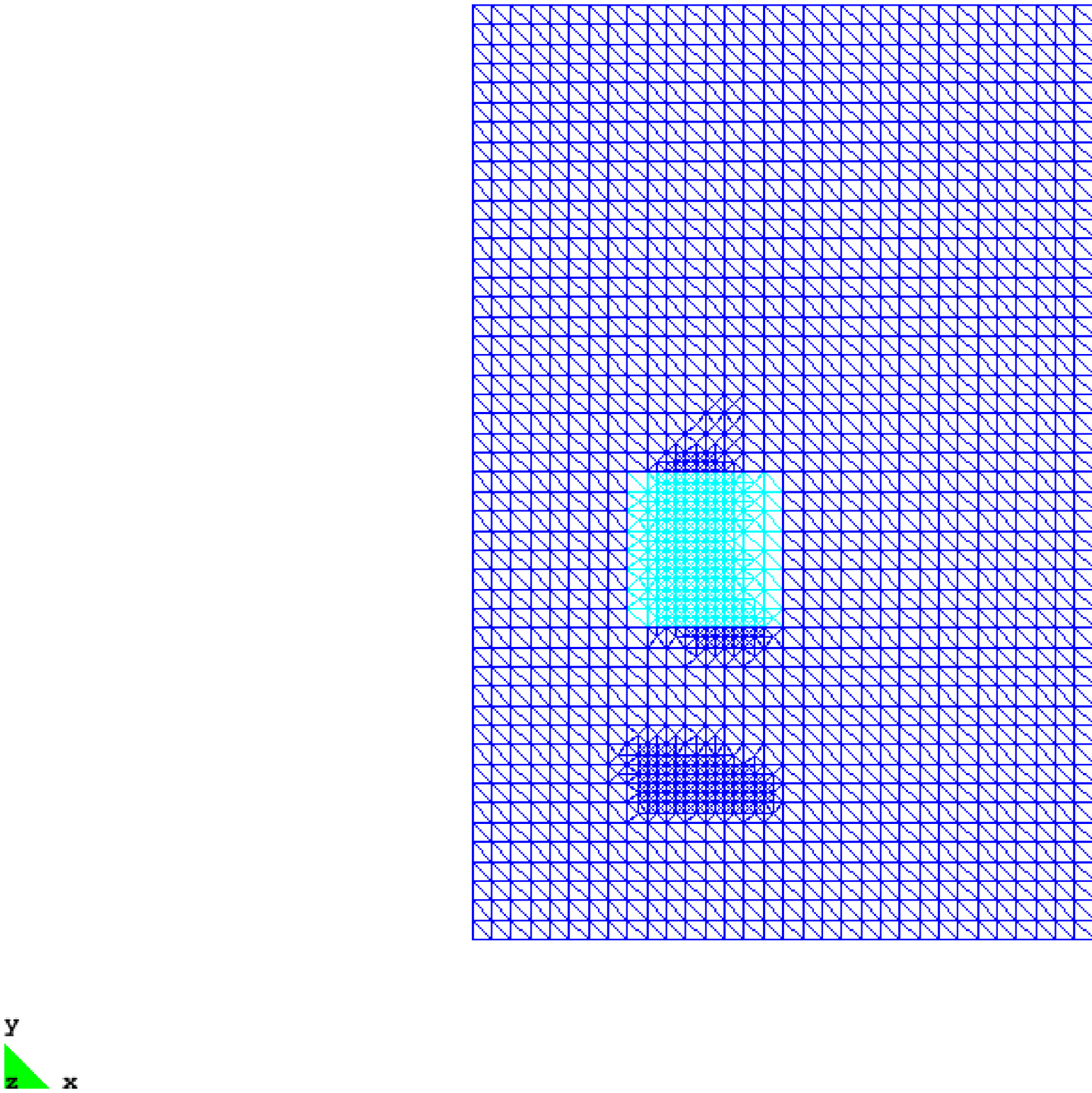}} \\
a) 4776 elements & b) 5272 elements & c) 6174 elements \\
{\includegraphics[scale=0.2,clip=]{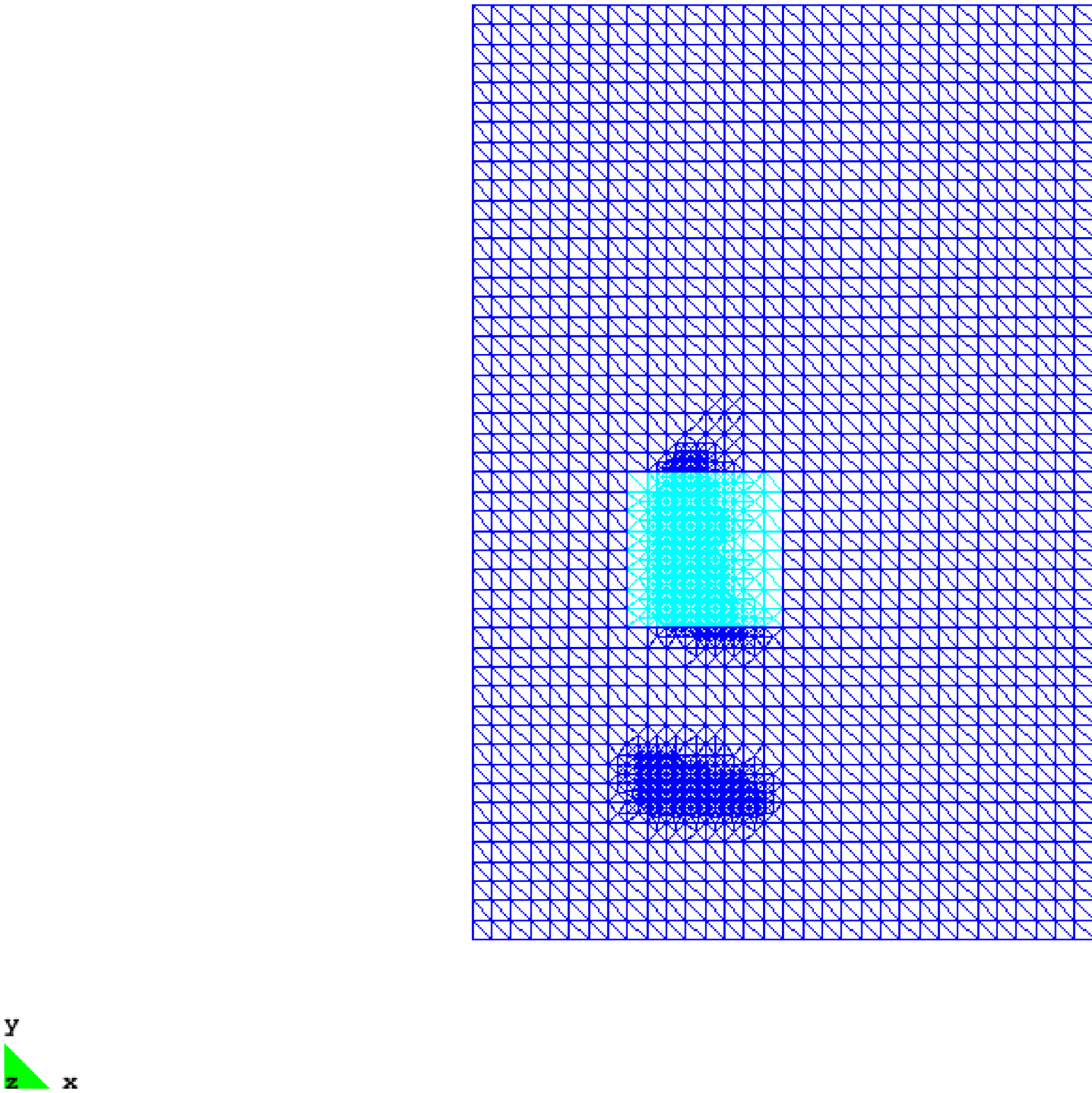}} &
{\includegraphics[scale=0.2,clip=]{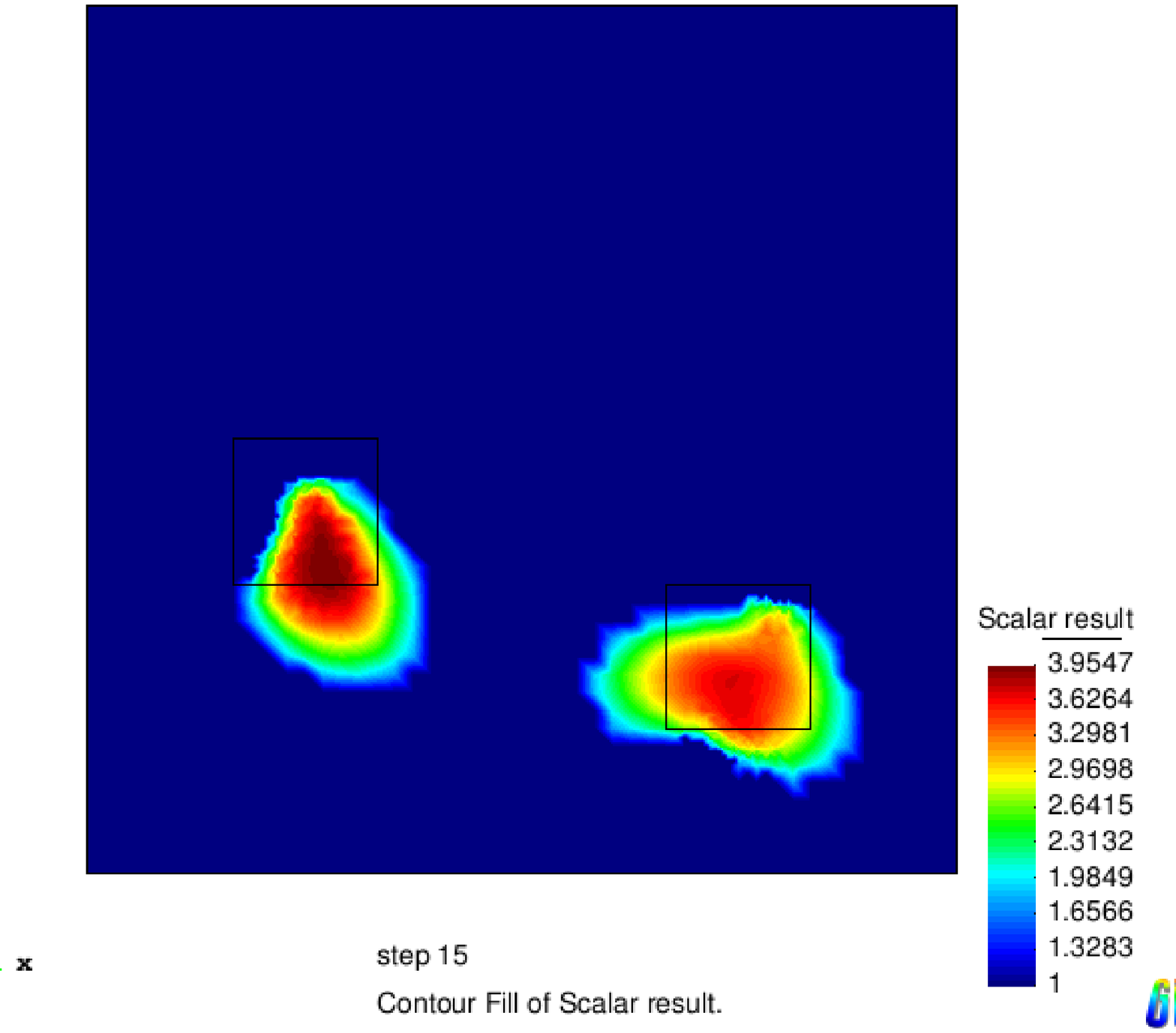}} &
{\includegraphics[scale=0.22,clip=]{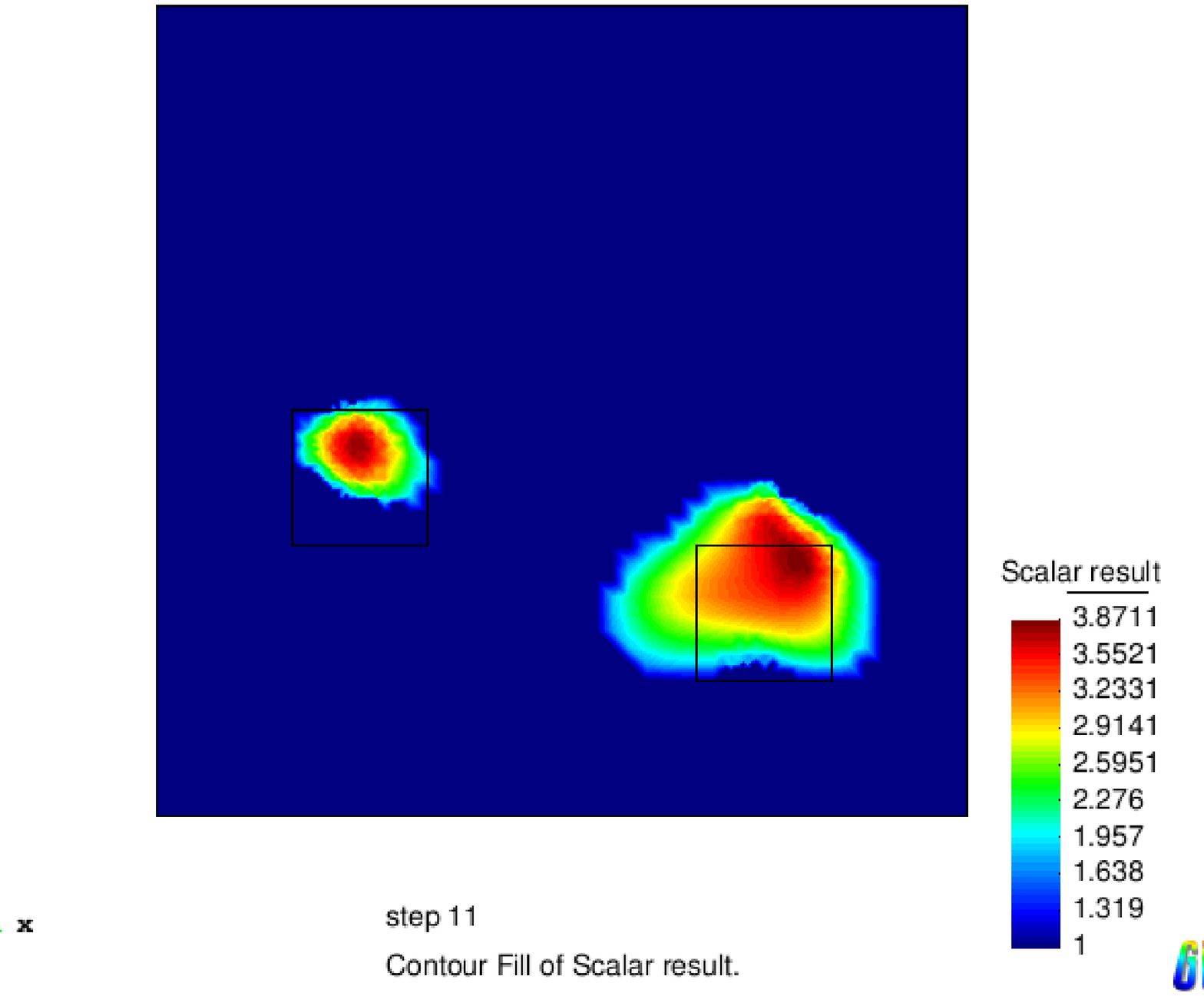}} \\
 d) 7682 elements & e)  $c_{4}(x) ,\max c_{4}(x) =3.9$ & f)   $c_{5}(x) ,\max c_{5}(x) =3.87$ \\
\end{tabular}
\caption{ Adaptively refined meshes (a)--(d) and finally reconstructed
  images (e) and (f) on 4-th and 5-th adaptively refined meshes,
  respectively.  On e) $\max c_{4}=3.9$ and on f) $\max c_{5}=3.87.$
  Reconstructed function on e) is obtained on the mesh presented on
  d). The mesh for the function on f) is not shown. Locations of both squares of Figure
  \ref{fig:Figure3}-a) as well as maximal values of the computed
  funtion $c_{glob}\left( x\right) $ in them are imaged
  accurately.}
\label{fig:Figure3}
\end{center}
\end{figure}

\begin{figure}[tbp]
\begin{center}
\begin{tabular}{cc}
 {\includegraphics[height=5cm,width=7cm,clip=]{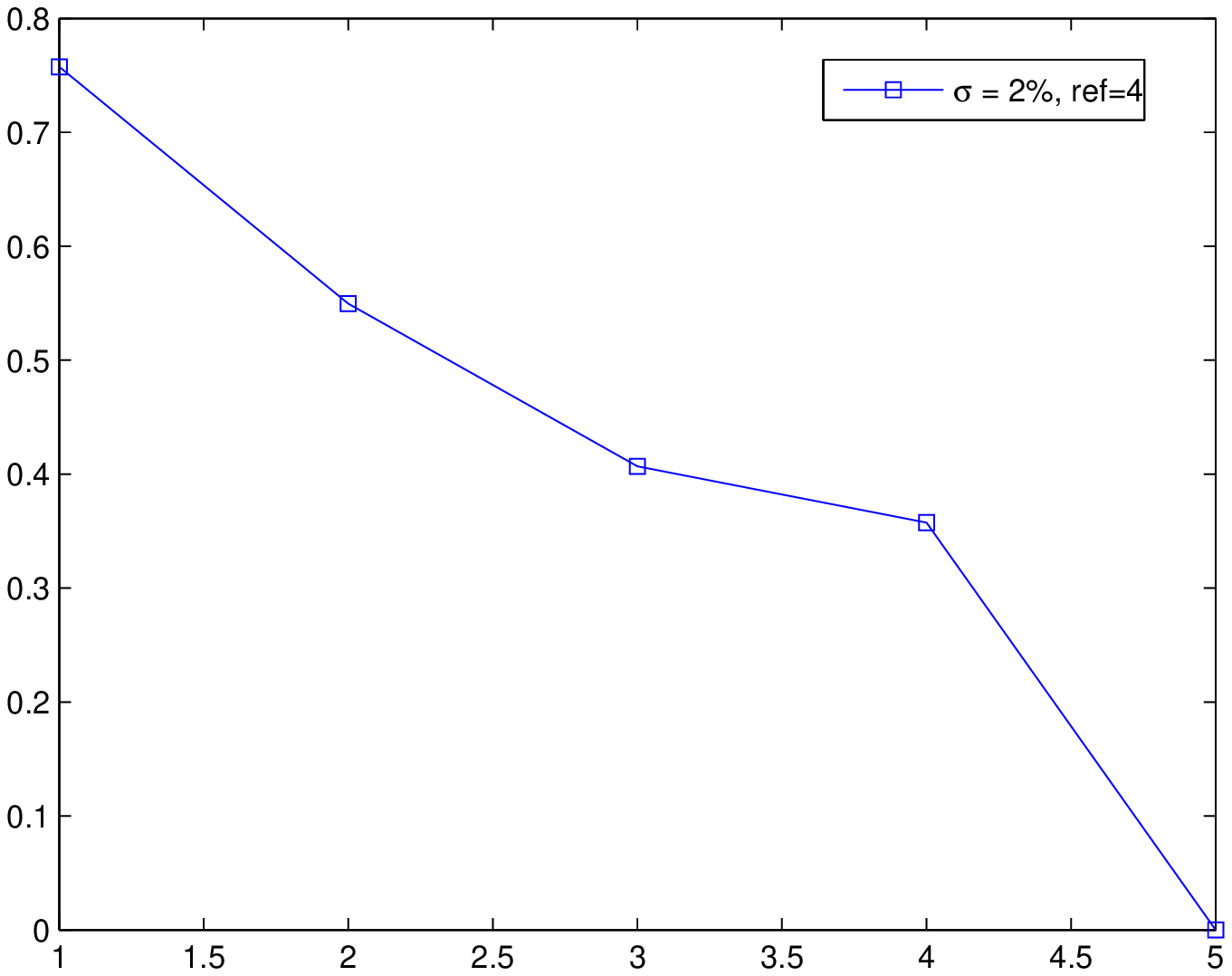}} &
{\includegraphics[height=5cm,width=7cm,clip=]{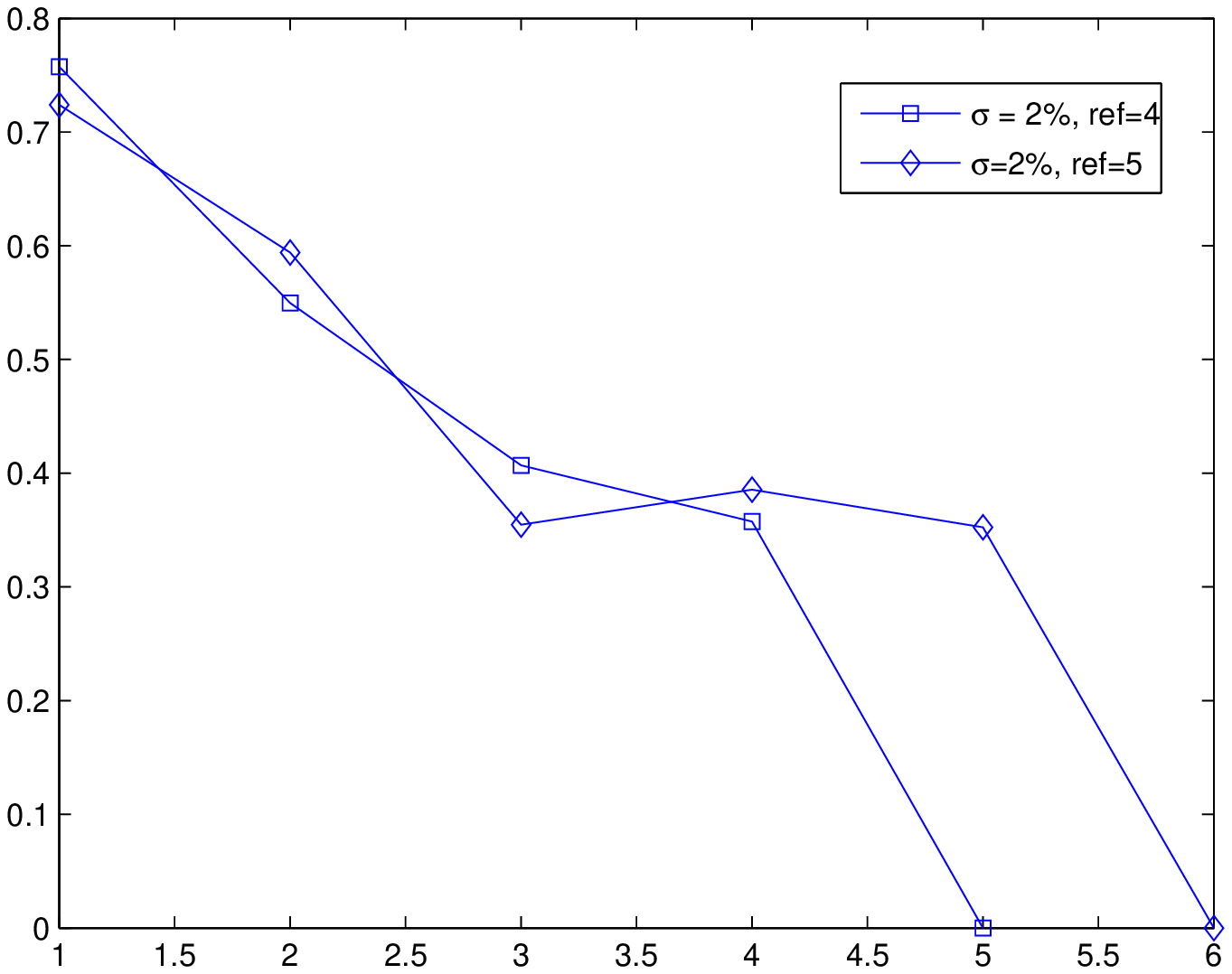}} \\
a)  & b) \\
\end{tabular}
\caption{ a) Computed relaxation property $||c_{n+1} -
  c_{\alpha}||_{L_2} \leq \eta_n ||c_{n} - c_{\alpha}||_{L_2}$ for the
  noise level $2\%$ in (\ref{noise}) and the regularization parameter
  $\alpha = 0.02$ in (\ref{6.11}). Here, $0 <\eta_n < 1$ is the small
  relaxation parameter obtained after $n$ mesh refinements. Here, we
  take $c_{\alpha}$ on the 4-th refined mesh shown on the Figure
  \ref{fig:Figure3}-d).  b) Comparison of the relaxation property
  $||c_{n+1} - c_{\alpha}||_{L_2} \leq \eta_n ||c_{n} -
  c_{\alpha}||_{L_2}$ when we take different functions $c_{\alpha}$:
  on the 4-th or on the 5-th refined mesh.}
\label{fig:Figure4}
\end{center}
\end{figure}

\emph{2. The adaptivity stage.} Since we have observed that $u\left(
x,T\right) \approx 0,$ we have not used the function $z_{\zeta }\left(
t\right) $ in our computations.
In this test we take the noise level $2\%$ in (\ref{noise}) and
  the regularization parameter $\alpha = 0.02$ in
  (\ref{6.11}).
% In this test we take the
%regularization parameter $\alpha = 0.02$.
 We now comment on the stopping
criterion for mesh refinements, which we use in numerical studies of
the adaptivity technique in this paper.  Let $c_{n}$ is the coefficient $c(x)$ calculated
after $n$ mesh refinements.  In Theorems 5.2-5.4, 6.5, 6.6 the
relaxation parameter $\eta$ is independent on the mesh refinement
number $n$. In practice, however, one should expect such dependence
$\eta:= \eta_n$. In this case the parameter $\eta$ of those theorems is
$\eta= \max(\eta_n)$.
Then because of the relaxation
property of Theorems 6.5, 6.6 as well as because of Remark 5.1, it is
anticipated that numbers $\eta_{n}$ decrease with the grow of $n$ until the
regularized solution $c_{\alpha \left( \delta \right) }$ is approximately
reached. However, nothing can be guaranteed about numbers $\eta_{n}$ as soon as
the regularized solution is reached. Hence, in our computations of the
adaptivity method we stopped mesh refinement process at such $n:=n_{0}$ that
$\eta_{n_{0}}> \eta_{n_{0}-1}.$ If $\eta_{n_{0}}\approx \eta_{n_{0}-1},$ then we took the
final solution $c_{final}:= c_{n_{0}}.$

Figure \ref{fig:Figure3}-e), f) represents the images obtained after 4
and 5 mesh refinements, respectivelly, as well as adaptive locally
refined meshes are presented on \ref{fig:Figure3}-a)-d). Comparing
with Figure \ref{fig:Figure1}-c), one can observe that locations of
both inclusions are imaged accurately. Recall that in each inclusion
of Figure \ref{fig:Figure1}-c) $c\left( x\right) =4$, see definition
for $c(x)$ in (\ref{Y}) shown also on Figure
\ref{fig:Figure2}-a). Therefore, maximal values of the function
$c\left( x\right) $ on Figures \ref{fig:Figure3}-e),f) are also
accurately imaged: the error does not exceed 3.5\%.

 Figure \ref{fig:Figure4} displays the graph of the dependence of the
 norm $\left\Vert c_{n}-c_{\alpha}\right\Vert _{L_{2}\left( \Omega
   \right) }$ from the mesh refinement number $n$.  By (\ref{6.22})
 and (\ref{6.23}) these norms should decay. Since we do not exactly
 know what the regularized solution $c_{\alpha }$ is, we have taken
 $c_{\alpha }:=c_{4}$ on Figure \ref{fig:Figure4}-a). On Figure
 \ref{fig:Figure4}-b) we have superimposed those graphs for $c_{\alpha
 }:=c_{4}$ and $c_{\alpha }:=c_{5}.$ One can observe that norms
 $\left\Vert c_{n}-c_{\alpha }\right\Vert $ decay in the case when
 $c_{\alpha}$ is taken on the 4-th refined mesh. At the same time we
 also observe, that the relaxation property (\ref{6.23}) is not
 fullfilled when we take $c_{\alpha}$ on the 5-th refined mesh since
 $\eta_3 > \eta_2$, see \ref{fig:Figure4}-b). Thus, we take the final
 reconstruction $c_{\alpha }:=c_{4},$ the function obtained after four
 (4) mesh refinements.

\textbf{Remark 8.1}. It is well known that\ imaging of locations of small
inclusions and maximal values of the function $c\left( x\right) $ in them is
of the primary interest in applications and it is more interesting than
imaging of slowly changing parts. Indeed, small inclusions can be explosives
\cite{KBK,KBKSNF}, tumors, etc..

\textbf{Remark 8.2.} The above stopping criterion for mesh refinements shows
that relaxation Theorems 6.5, 6.6 are quite useful for computations.

\subsection{Experimental data}

\label{sec:8.2}

Experimental studies were described in detail in \cite{BK4,KFBPS} as well as
in Chapter 5 of \cite{BK1}. Hence, we omit many details here. We point out
that the main difficulty was a \emph{huge misfit} between computationally
simulated and experimental data. The latter was the case even for the free
space data: the analytic solution predicted by Maxwell equations was
radically different from the experimentally measured curves. This can be
explained by unknown nonlinear processes in both transmitters and detectors.
The same was observed for the backscattering data collected in the field,
see \cite{KBKSNF} and section 6.9 of \cite{BK1}. To handle this misfit, a
new data pre-processing procedure was applied. This procedure has immersed
experimental data in computationally simulated ones, see Figures 4 in \cite%
{KBKSNF} and Figures 5.3 in \cite{BK1}. Naturally, this procedure has
introduced a significant modeling noise in already noisy data. Nevertheless,
computational results were very accurate ones, which speaks well for the
robustness of our reconstruction method. The first stage of our two-stage
numerical procedure was working with \emph{blind} data (unlike the second
stage). Therefore, results of at least the first stage were unbiased.

\begin{figure}[tbp]
\centerline{\includegraphics[width=0.75\textwidth]{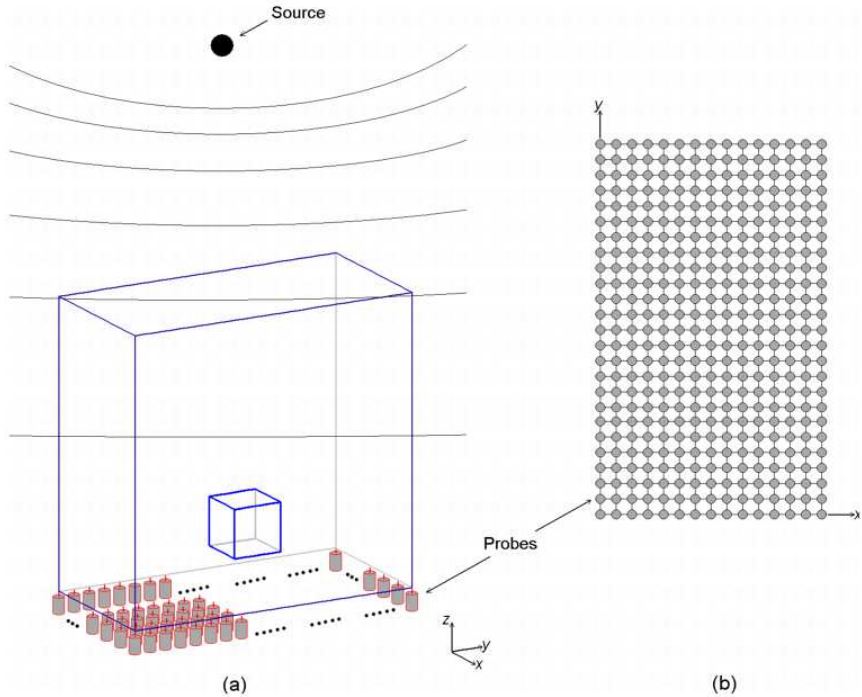}}
\caption{\emph{Schematic diagram of data collection. Original source:
      M.~V.~Klibanov, M.~A.~Fiddy, L.~Beilina, N.~Pantong and
      J.~Schenk, Picosecond scale experimental verification of a
      globally convergent numerical method for a coefficient inverse
      problem, \emph{Inverse Problems}, 26, 045003,
      doi:10.1088/0266-5611/26/4/045003, 2010. \copyright IOP
      Publishing.  Reprinted with permission. }
}
\label{fig:Figure5}
\end{figure}

The data collection scheme is displayed on Figure \ref{fig:Figure5}. A single source of
electric wave field emits pulse for only one component of the electric
field, two other components were not emitted. The prism is our computational
domain $\Omega .$ The outcome time resolved signal was measured at many
detectors located on the bottom side of the prism. The same component of the
electric field was measured as the one emitted. Since we have not measured
that signal at the rest $\partial _{1}\Omega $ of $\partial \Omega ,$ we
have prescribed to $\partial _{1}\Omega $ the same boundary conditions as
ones for the uniform medium with the dielectric constant $\varepsilon
_{r}\equiv 1.$ The prism $\Omega $ \ is filled with a dielectric material
with the dielectric constant $\varepsilon _{r}\approx 1,$ i.e. almost the
same as in the air. We point out, however, that when using the first stage
of our two-stage numerical procedure, we did not use any knowledge of the
dielectric constant of this prism. We have only used the fact that $%
\varepsilon _{r}=1$ outside of this prism, see (\ref{6.1}).

We have placed one dielectric inclusion inside of this prism. Inclusions
were two wooden cubes, which we call below \textquotedblleft Cube 1" and
\textquotedblleft Cube 2". Sizes of their sides were 4 cm for Cube 1 and 6
cm for Cube 2. Note that only refractive indices $n=\sqrt{\varepsilon _{r}}$
rather than dielectric constants can be measured directly in experiments.
The goal of the first stage was to reconstruct the refractive index of the
inclusion and its location. The goal of the second stage was to reconstruct
all three components of inclusions: refractive indices, shapes and
locations. Since only one component of the electric field was measured, we
have modeled the wave propagation process via the problem (\ref{8.1}) with $%
\varepsilon _{r}\left( x\right) :=c\left( x\right) $, where the domain $%
G\subset \mathbb{R}^{3}$ was a prism, which was bigger than the prism $%
\Omega ,$ see (5.8) and section 5.4 in \cite{BK1} for this domain. The
function $f\left( t\right) $ in (\ref{8.1}) was
\begin{equation*}
f\left( t\right) =\left\{
\begin{array}{c}
\sin \left( \omega t\right) ,t\in \left( 0,2\pi /\omega \right) , \\
0,t>2\pi /\omega ,%
\end{array}%
\right.
\end{equation*}%
where $\omega =14$ for Cube 1 and $\omega =7$ for Cube 2 (see page 329 of
\cite{BK1} and page 26 of \cite{BK4} for $\omega $). It was only later,
after the first author has conducted numerical simulations for solving the
Maxwell equations \cite{Beilina2}, when we have realized that the choice of
modeling by one PDE only was well justified.
%Indeed, it was demonstrated
%computationally in \cite{Beilina2} that the component of the electric field,
%which was initialized, dominates two other components.
In our experiments,
Cubes 1 and 2 were placed total in six different positions.

\begin{table}[tp]
\begin{center}
\begin{tabular}{|c|c|c|c|}
\hline\\
Case number & Computed $n$ & Directly measured $n$ & Computational error \\
\hline
1 (Cube 1) & 1.97 & 2.07 & 5\% \\
2 (Cube 1) & 2 & 2.07 & 3.4\% \\
3 (Cube 1) & 2.16 & 2.07 & 4.3\% \\
4 (Cube 1) & 2.19 & 2.07 & 5.8\% \\
5 (Cube 2) & 1.73 & 1.71 & 1.2\% \\
6 (Cube 2) & 1.79 & 1.71 & 4.7\% \\
\hline
\end{tabular}
\end{center}
\caption{Blindly computed and directly measured refractive indices $%
n$ by the first stage of our two-stage numerical procedure. The error in
direct measurements was 11\% for cases 1-4 (Cube 1) and 3.5\% for cases 5,6
(Cube 2).}
\label{tab:Table1}
\end{table}

Table \ref{tab:Table1}
summarizes results of blind study of the first stage of our two-stage
numerical procedure. Because of the blind test requirement, direct
measurements of refractive indices were performed by the conventional
so-called \textquotedblleft waveguide method" \cite{RMC}\emph{\ only after}
computations of the first stage were done. Next, computational results were
compared with measured ones. One can see that we had only a few percent
difference with \emph{a posteriori} directly measured refractive indices of
both cubes. Furthermore, in five out of six cases this error was even less
than the error in direct measurements.

\begin{figure}[tbp]
\begin{center}
\begin{tabular}{cc}
 {\includegraphics[scale=0.3, angle=-90,clip=]{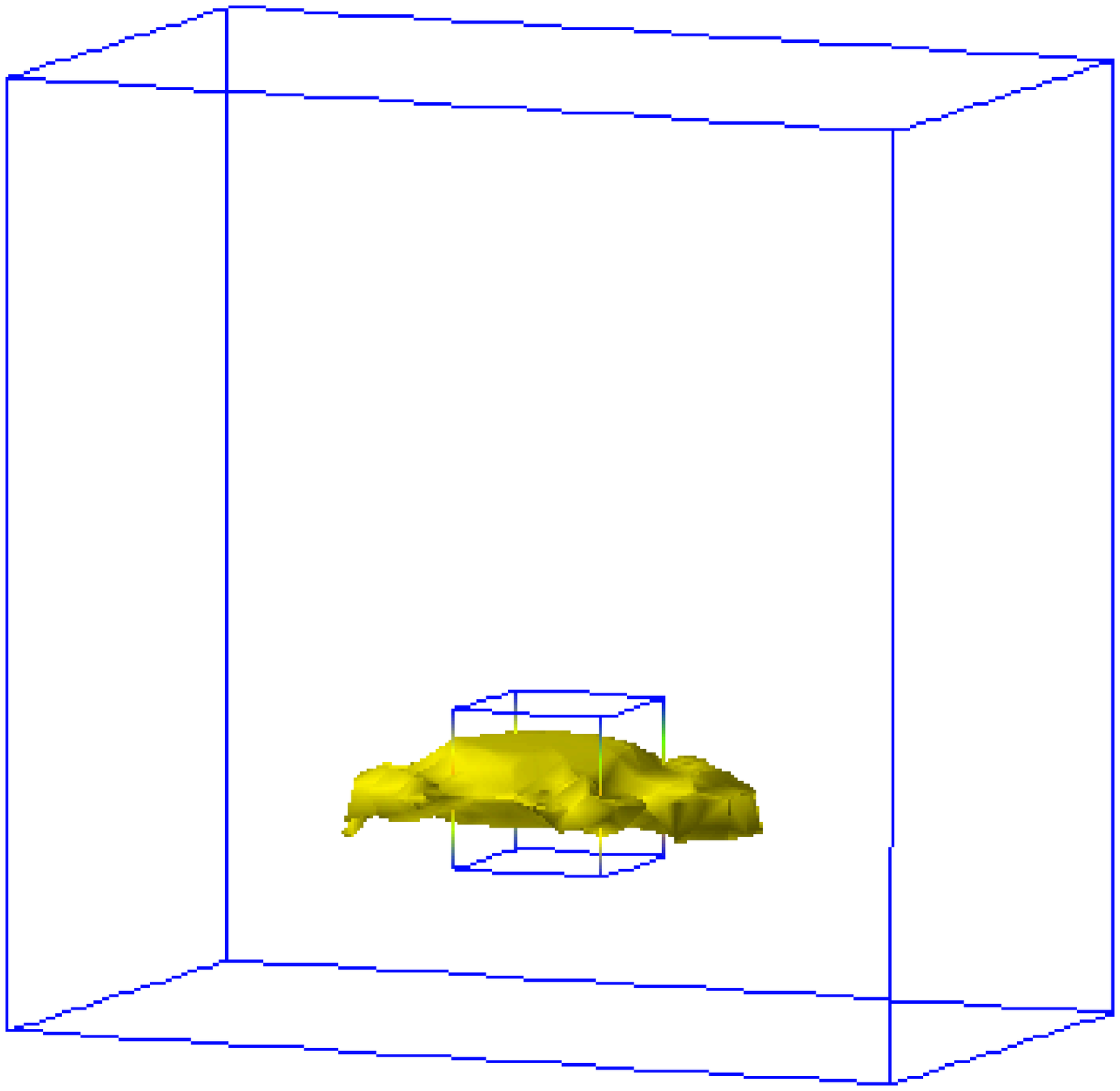}} &
 {\includegraphics[scale=0.27, angle=-90,clip=]{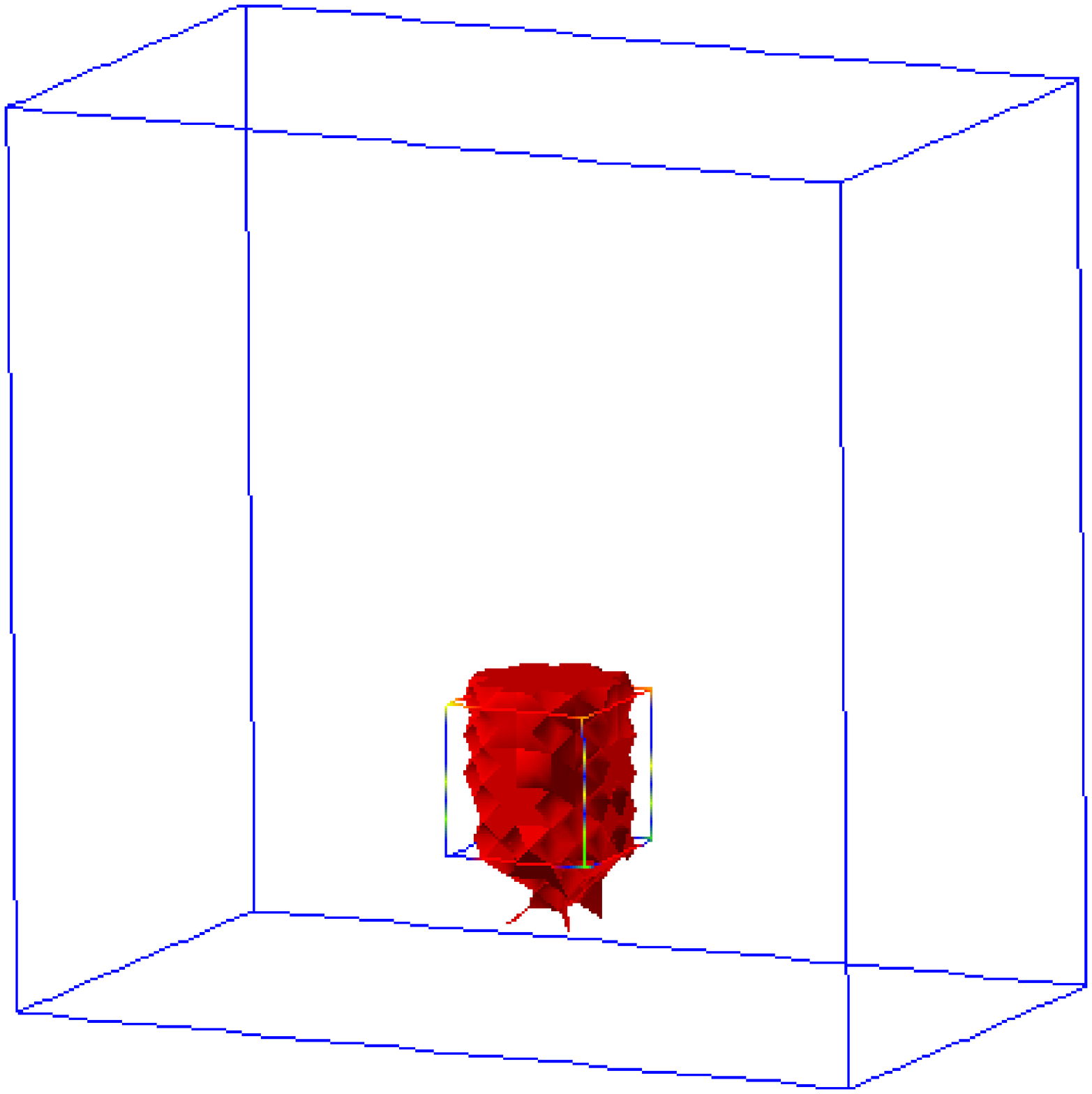}} \\
a)  $\max n_{glob}=\max \sqrt{\varepsilon _{r,glob}}=\max \sqrt{c_{glob}}=1.97$ &
  b) $\max n=\max \sqrt{\varepsilon _{r}}=\max \sqrt{c}=2.05$ \\
\end{tabular}
\end{center}
\caption{ Case 1 of Table  \ref{tab:Table1} was tested by the two-stage numerical
procedure. a) The computational result of the first stage. Both location of
the inclusion and refractive index $n_{glob}=1.97$ are accurately reconstructed.
However, the shape of the inclusion is not reconstructed accurately. b) The
computational result of the second (refinement) stage. All three components
of the inclusion are very accurately reconstructed: refractive index,
location and shape. Also, values of the function $\varepsilon
_{r}\left( x\right) :=c\left( x\right) =1$ outside of the imaged inclusion
are computed very accurately. Original source: L. Beilina and M.V. Klibanov,
        Reconstruction of dielectrics from experimental data via a
        hybrid globally convergent/adaptive inverse algorithm,
        \emph{Inverse Problems}, 26, 125009,
        doi:10.1088/0266-5611/26/12/125009, 2010. \copyright IOP
        Publishing. Reprinted with permission.
}
\label{fig:Figure6}
\end{figure}

We now focus on the results which we have obtained on the second stage of
our two-stage numerical procedure when applying the adaptivity.

\textbf{Test 2. The two stage numerical procedure for Case 1 of Table  \ref{tab:Table1}. }
Figure \ref{fig:Figure6}-a) displays\ the result of the first stage of the two-stage
numerical procedure. One can see that although the refractive index $n=1.97$
and location of the inclusion are accurately calculated, the shape is
inaccurate. The image of Figure \ref{fig:Figure6}-a) was taken as the starting point for the
adaptivity technique for refinement. The result of the second stage is
presented on Figure \ref{fig:Figure6}-b). One can see that all three components of the
inclusion are accurately reconstructed. In addition, the values of the
function $\varepsilon _{r}\left( x\right) :=c\left( x\right) =1$ outside of
the imaged inclusion are also accurately computed.

\begin{figure}[tbp]
\begin{center}
\begin{tabular}{cc}
 {\includegraphics[scale=0.3, angle=-90,clip=]{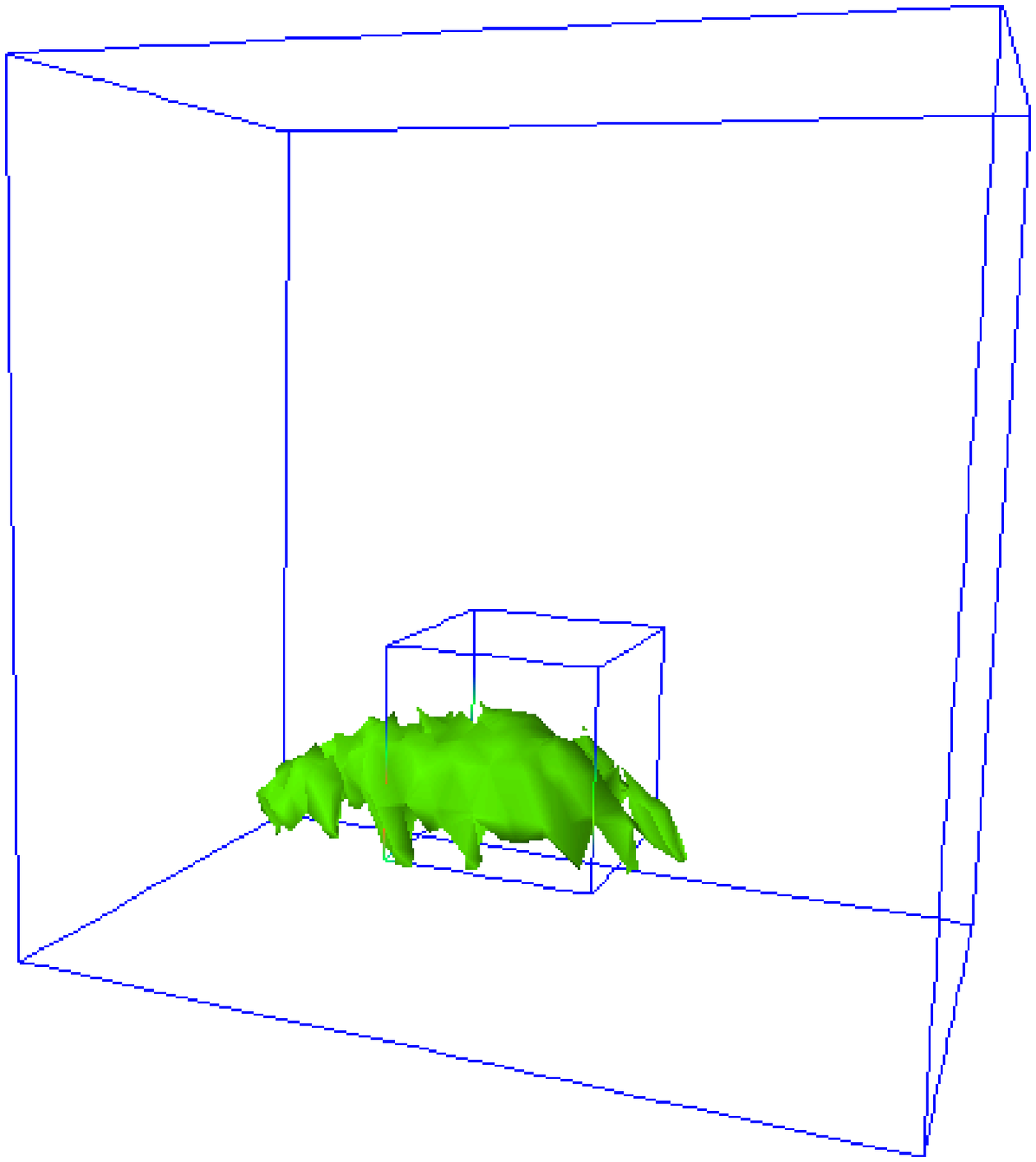}} &
 {\includegraphics[scale=0.27, angle=-90,clip=]{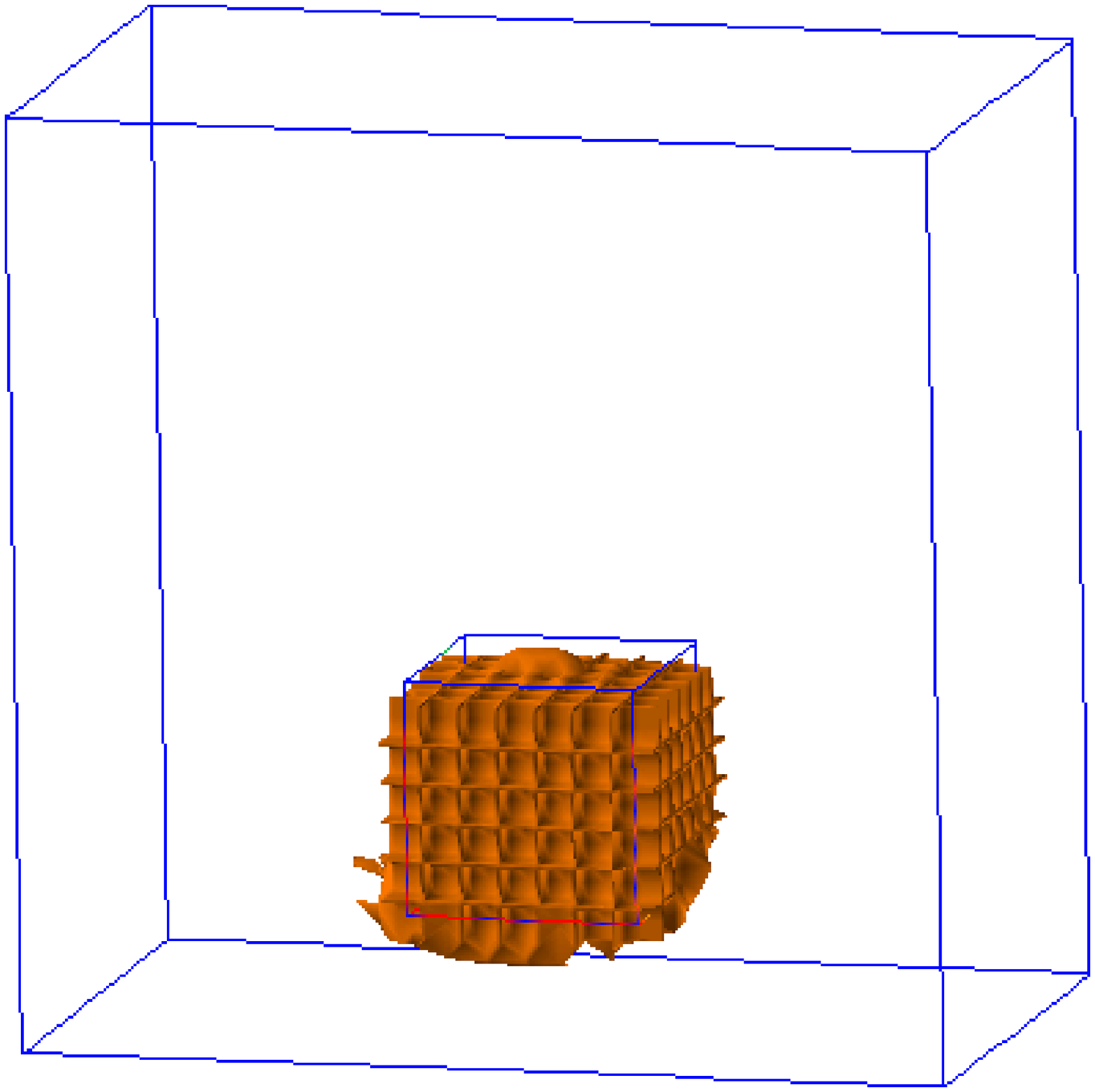}} \\
a)  $\max n_{glob}=\max \sqrt{\varepsilon _{r,glob}}=\max \sqrt{c_{glob}}=1.79$  &
  b)  $\max n=\max \sqrt{\varepsilon _{r}} =\max \sqrt{c}=1.73$ \\
\end{tabular}
\end{center}
\caption{Case 6 of Table \ref{tab:Table1} was tested for the
  two-stage numerical procedure. a) The computational result of the
  first stage. Both location of the inclusion and refractive index
  $n_{glob}=1.79$ are accurately reconstructed. However, the shape of the
  inclusion is not reconstructed accurately. b) The computational
  result of the second (refinement) stage. All three components of the
  inclusion are accurately reconstructed: location, refractive index,
  and shape\textbf{. }In addition, values of the function $\varepsilon
  _{r}\left( x\right) :=c\left( x\right) =1 $ outside of the imaged
  inclusion are computed accurately. Original source: L. Beilina and
  M.V. Klibanov, Reconstruction of dielectrics from experimental data
  via a hybrid globally convergent/adaptive inverse algorithm,
  \emph{Inverse Problems}, 26, 125009,
  doi:10.1088/0266-5611/26/12/125009, 2010. \copyright IOP
  Publishing. Reprinted with permission.  }
\label{fig:Figure7}
\end{figure}

\textbf{Test 3. The two stage numerical procedure for Case 6 of Table
  \ref{tab:Table1}.}  Figures \ref{fig:Figure7}-a) and
\ref{fig:Figure7}-b) display computational results for first and
second stages, respectively. The rest of comments are the same as ones
for Test 2.  Note that the shape is now reconstructed better than in
Test 2. This can be heuristically explained as follows. The wavelength
of our electromagnetic wave  was $\mu =3$ cm. Thus, the size of the side of Cube 1 is
4 cm=1.33$\mu .$ One the other hand, the size of the side of Cube 2 is
6 cm=2$\mu ,$ which is larger.

\vspace{1cm}

\begin{center}
\textbf{Acknowledgments}
\end{center}

This research was supported by US Army Research Laboratory and US Army
Research Office grant W911NF-11-1-0399, the Swedish Research Council, the
Swedish Foundation for Strategic Research (SSF) in Gothenburg Mathematical
Modelling Centre (GMMC) and by the Swedish Institute, Visby Program.

\end{document}